\documentclass[aps,twocolumn,showpacs,prx,superscriptaddress,preprintnumbers,amsmath,amssymb]{revtex4-1}

\ifx\pdfoutput\undefined
\usepackage[dvipdfmx]{graphicx}
\usepackage[dvipdfmx]{hyperref}
\else
\usepackage{graphicx}
\usepackage{hyperref}
\fi
\usepackage[usenames]{xcolor}
\usepackage{dcolumn}% Align table columns on decimal point
\usepackage{bm}% bold math
\usepackage{ulem}

\begin{document}

\preprint{APS/123-QED}

\title{
Majorana dynamical mean-field study of spin dynamics at finite temperatures in the honeycomb Kitaev model
}

\author{Junki Yoshitake}
\affiliation{Department of Applied Physics, University of Tokyo, Bunkyo, Tokyo 113-8656, Japan}
\author{Joji Nasu}
\affiliation{Department of Physics, Tokyo Institute of Technology, Meguro, Tokyo 152-8551, Japan}
\author{Yasuyuki Kato}
\affiliation{Department of Applied Physics, University of Tokyo, Bunkyo, Tokyo 113-8656, Japan}
\author{Yukitoshi Motome}
\affiliation{Department of Applied Physics, University of Tokyo, Bunkyo, Tokyo 113-8656, Japan}


%\author{}
%\homepage{}
%\affiliation{
% with \\
%}%
%\affiliation{
%\affiliation{%


\date{\today}% It is always \today, today,

\begin{abstract}
A prominent feature of quantum spin liquids is fractionalization of the spin degree of freedom. 
Fractionalized excitations have their own dynamics in different energy scales, and hence,
affect finite-temperature ($T$) properties in a peculiar manner
even in the paramagnetic state harboring the quantum spin liquid state.
We here present a comprehensive theoretical study of the spin dynamics in a wide $T$ range for the Kitaev model on a honeycomb lattice, whose ground state is such a quantum spin liquid. 
In this model, the fractionalization occurs to break up quantum spins into itinerant matter fermions and localized gauge fluxes, which results in two crossovers at very different $T$ scales. 
Extending the previous study for the isotropic coupling case [J. Yoshitake, J. Nasu, and Y. Motome, Phys. Rev. Lett. {\textbf 117}, 157203 (2016)], we calculate the dynamical spin structure factor $S(\bf{q},\omega)$, the NMR relaxation rate $1/T_1$, and the magnetic susceptibility $\chi$ while changing the anisotropy in the exchange coupling constants, by using the dynamical mean-field theory based on a Majorana fermion representation.
We describe the details of the methodology including the continuous-time quantum Monte Carlo method for computing dynamical spin correlations and the maximum entropy method for analytic continuation.
We confirm that the combined method provides accurate results in a wide $T$ range including the region where the spins are fractionalized.
We find that also in the anisotropic cases the system exhibits peculiar behaviors below the high-$T$ crossover whose temperature is comparable to the average of the exchange constants: $S(\bf{q},\omega)$ shows an inelastic response at the energy scale of the averaged exchange constant, $1/T_1$ continues to grow even though the equal-time spin correlations are saturated and almost $T$ independent, and $\chi$ deviates from the Curie-Weiss behavior.
In particular, when the exchange interaction in one direction is stronger than the other two, the dynamical quantities exhibit qualitatively different $T$ dependences from the isotropic case at low $T$, reflecting the opposite parity between the flux-free ground state and the flux-excited state, and a larger energy cost for flipping a spin in the strong interaction direction.
On the other hand, when the exchange anisotropy is in the opposite way, the results are qualitatively similar to those in the isotropic case.
All these behaviors manifest the spin fractionalization in the paramagnetic region. 
Among them, the dichotomy between the static and dynamical spin correlations is unusual behavior hardly seen in conventional magnets.
We discuss the relation between the dichotomy and the spatial configuration of gauge fluxes.
Our results could stimulate further experimental and theoretical analyses of candidate materials for the Kitaev quantum spin liquids.

%\begin{description}
%\end{description}
\end{abstract}

%\pacs{71.10.Fd, 71.27.+a, 75.10.-b}
\maketitle

\section{Introduction}

Quantum many-body systems show various intriguing phenomena which cannot be understood as an assembly of independent particles.
One of such phenomena is fractionalization, in which the fundamental degree of freedom in the system is fractionalized into several quasiparticles.
A well-known example of such fractionalization is the fractional quantum Hall effect,
in which the Hall conductance shows plateaus at fractional values of $e^2/h$ ($e$ is the elementary charge and $h$ is the Planck constant)~\cite{Tsui1982,Stormer1999}.
In this case, the quasiparticles carry a fractional value of the elementary charge, as a collective excitation of the elementary particle, electron.
This is fractionalization of charge degree of freedom. 
On the other hand,  another degree of freedom of electrons, spin, can also be fractionalized.
Such a peculiar phenomenon has been argued for quantum many-body states in insulating magnets, e.g., a quantum spin liquid (QSL) state.

QSLs are the magnetic states which preserve all the symmetries in the high-temperature($T$) paramagnet even in the ground state and evade a description by conventional local order parameters.
A typical example of QSLs is the resonating valence bond~(RVB) state, proposed by P. W. Anderson~\cite{Anderson1973}.
The RVB state is a superposition of valence bond states (direct products of spin singlet dimers), which does not break either time reversal or translational symmetry.
In the RVB state, the spin degree of freedom is fractionalized: the system exhibits two different types of elementary excitations called spinon and vison~\cite{Kivelson1987,Senthil2000}. 
Spinon is a particlelike excitation carrying no charge but spin $S=1/2$.
Meanwhile, vison is a topological excitation characterized by the parity of crossing singlet pairs with its trace.
Another example of QSLs is found in quantum spin ice systems, in which peculiar excitations are assumed to be magnetic monopoles, electric gauge charges, and artificial photons resulting from fractionalization of the spin degree of freedom~\cite{Hermele2004,Castelnovo2008}.

Among theoretical models for QSLs, the Kitaev model has attracted growing interest, as it realizes the fractionalization of quantum spins in a canonical form~\cite{Kitaev2006}.
The Kitaev model is a localized spin $S=1/2$ model defined on a two-dimensional honeycomb lattice with bond-dependent anisotropic interactions (see Sec.~\ref{subsec:Majorana_rep}).
In this model, the ground state is exactly obtained as a QSL, in which quantum spins $S=1/2$ are fractionalized into itinerant Majorana fermions and localized gauge fluxes.
The fractionalization affects the thermal and dynamical properties in this model.
For instance, the different energy scales between the fractionalized excitations appear as two crossovers at largely different $T$ scales; in each crossover, itinerant Majorana fermions and localized gauge fluxes release their entropy, a half of $\log 2$ per site~\cite{Nasu2014,Nasu2015}. 
Also in the ground state, the dynamical spin structure factor $S(\mathbf{q}, \omega)$ shows a gap due to the flux excitation and strong incoherent spectra from the composite excitations between itinerant Majorana fermions and localized gauge fluxes~\cite{Knolle2014}.
Such incoherent spectra were indeed observed in recent inelestic neutron scattering experiments for a candidate for the Kitaev QSL, $\alpha$-RuCl$_3$~\cite{Banerjee2016,Do_preprint}.
The magnetic Raman scattering spectra also shows a broad continuum dominated by the itinerant Majorana fermions, in marked contrast to conventional insulating magnets~\cite{Knolle2014b}. 
Such a broad continuum was experimentally observed also in $\alpha$-RuCl$_3$~\cite{Sandilands2015}.
Furthermore, the $T$ dependence of the incoherent response was theoretically analyzed and identified as the fermionic excitations emergent from the spin fractionalization~\cite{Nasu2016, Glamazda2016}.

In the previous study, the authors have studied dynamical properties of the Kitaev model at finite $T$ by a newly developed numerical technique, the Majorana dynamical mean-field method~\cite{Yoshitake2016}. 
Indications of the spin fractionalization were identified in the $T$ dependences of $S(\mathbf{q}, \omega)$, the relaxation rate in the nuclear magnetic resonance (NMR), $1/T_1$, and the magnetic susceptibility $\chi$.
In the previous study, however, the results were limited to the case with the isotropic exchange constants, despite the anisotropy existing in the Kitaev candidate materials~\cite{Johnson2015, Yamaji2014}.
In the present paper, to complete the analysis, we present the numerical results of the dynamical quantities for anisotropic cases.
We also provide the comprehensive description of the theoretical method, including the details of the cluster dynamical mean-field theory (CDMFT), the continuous-time quantum Monte Carlo (CTQMC) as a solver of the impurity problem to calculate the dynamical spin correlations, and the maximum entropy method (MEM) for the analytic continuation.
We discuss a prominent feature proximate to the QSL, i.e., the dichotomy between static and dynamical spin correlations, from the viewpoint of the fractionalization of spins. 

The paper is organized as follows. 
In Sec.~\ref{sec:ModelAndMethod}, after introducing the Kitaev model and its Majorana fermion representation, we describe the details of the CDMFT, CTQMC, and MEM.
In Sec.~\ref{sec:Result}, we show the numerical results for $S(\mathbf{q}, \omega)$, $1/T_1$, and $\chi$ while changing the anisotropy in the exchange constants. 
In Sec.~\ref{sec:Discussion}, we discuss the dichotomy between the static and dynamical spin correlations by comparing the $T$ dependences for the uniform and random flux configurations.
The cluster-size dependence in the CDMFT is examined in Appendix~\ref{app:SizeDep}.
The accuracy of MEM is also examined in Appendix~\ref{app:MEM} in the one-dimensional limit where the dynamical properties can be calculated without analytic continuation.
We also show the $T$ and $\omega$ dependence of spin correlations and the $T$ dependence of the Korringa ratio in Appendix~\ref{app:S(w,T)} and \ref{app:Korringa}, respectively.

\section{Model and method}
\label{sec:ModelAndMethod}
In this section, we describe the details of the methods used in the present study, the Majorana CDMFT and CTQMC methods. 
After introducing the Majorana fermion representation of the Kitaev model in Sec.~\ref{subsec:Majorana_rep}, we describe the framework of the Majorana CDMFT in Sec.~\ref{subsec:CDMFT}, in which the impurity problem is solved exactly. 
In Sec.~\ref{subsec:CTQMC}, we introduce the CTQMC method which is applied to the converged solutions obtained by the Majorana CDMFT for calculating dynamical spin correlations. 
We also touch on the MEM used for obtaining the dynamical spin correlations as functions of real frequency from those of imaginary time in Sec.~\ref{subsec:MEM}.

\subsection{Kitaev model and the Majorana fermion representation}
\label{subsec:Majorana_rep}

\begin{figure}[t]
    \includegraphics[width=\columnwidth,clip]{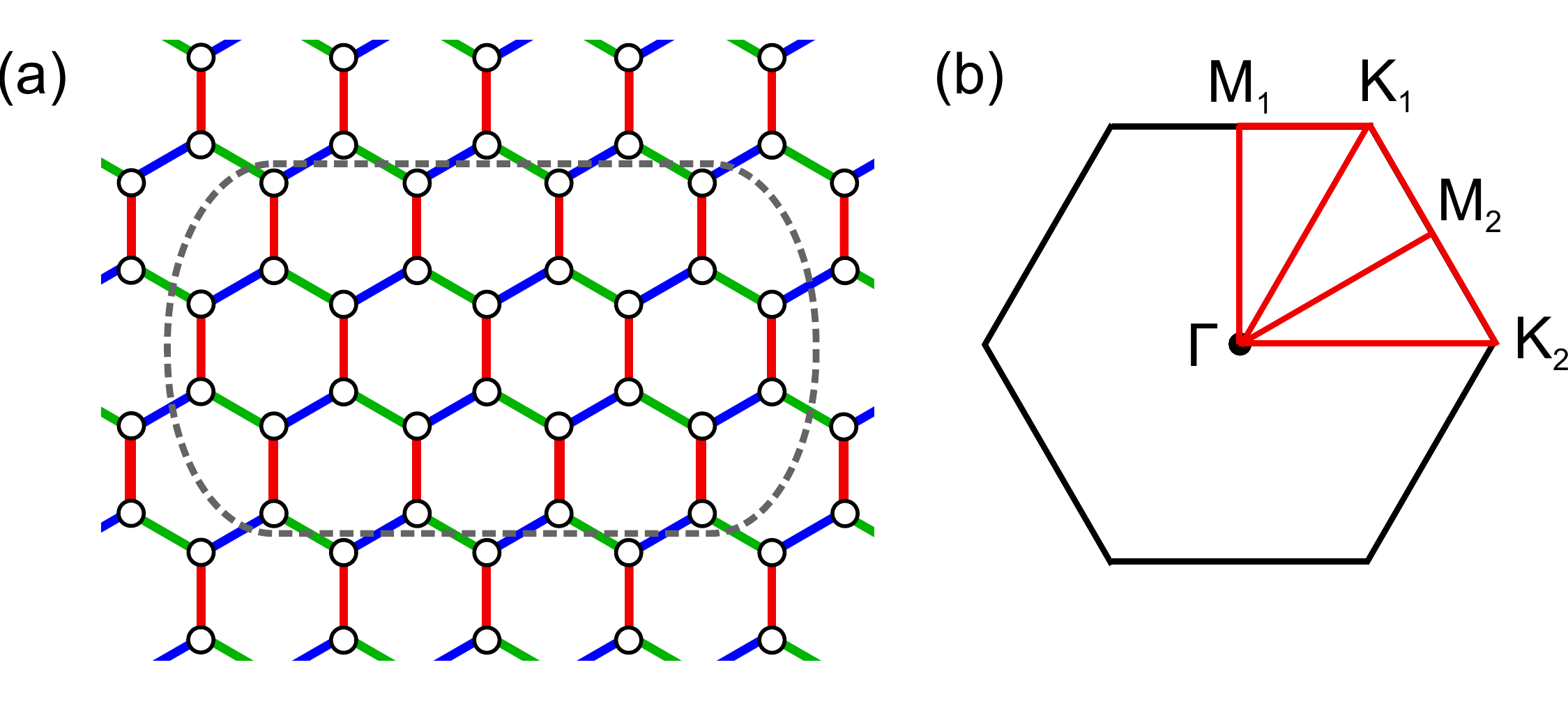}
    \caption{ \label{fig:fig1}
    (a) Schematic picture of the Kitaev model on the honeycomb lattice.
    The blue, green, and red bonds represent the $p=x,y$, and $z$ bonds in Eq.~(\ref{eq:Hamiltonian0}), respectively.
    The dashed oval represents the 26-site cluster used in the CDMFT calculations.
    (b) The first Brillouin zone (black hexagon) and the symmetric lines (red lines) used in Figs.~\ref{fig:fig3} and \ref{fig:fig4}.
    }
\end{figure}

We consider the Kitaev model on a honeycomb lattice, whose Hamiltonian is given by~\cite{Kitaev2006}
\begin{align}
\mathcal{H} = 
-\sum_p J_p \sum_{\langle j,j'\rangle_p} S_j^p S_{j'}^p,
\label{eq:Hamiltonian0}
\end{align}
where $p=x$, $y$, and $z$, and the sum of ${\langle j,j' \rangle}_p$ is taken for the nearest-neighbor (NN) sites on three inequivalent bonds of the honeycomb lattice, as indicated in Fig.~\ref{fig:fig1}(a); 
$S_{j}^p$ is the $p$ component of the $S=1/2$ spin at site $j$.
Hereafter, we denote the average of $J_p$ as $J$ and set the energy scale as $\sum_p |J_p| = 3$, i.e., $J=1$, and parametrize the anisotropy of the exchange coupling constants as $J_x=J_y=\pm\alpha$ and $J_z=\pm(3-2\alpha)$, where $+$ and $-$ correspond to the ferromagnetic (FM) and antiferromagnetic (AFM) cases, respectively. 
We note that the FM and AFM cases are connected through unitary transformations~\cite{Kitaev2006}.

As shown by Kitaev~\cite{Kitaev2006}, the model is soluble and the exact ground state is obtained as a QSL. 
The spin correlations are extremely short-ranged: $\langle S_j^p S_{j'}^p \rangle$ are nonzero only for the NN sites $j,j'$ on the $p$ bonds as well as the same site $j=j'$~\cite{Baskaran2007}. 
Hereafter, we denote the NN correlations as $\langle S_j^p S_{j'}^p \rangle_{\rm{NN}}$. 
There are two types of QSL phases depending on the anisotropy in $J_p$: one is a gapless QSL realized in the region with $0.75\leq \alpha \leq 1.5$ including the isotropic point $\alpha=1$ ($J_x=J_y=J_z=\pm1$), while the other is gapful for $0\leq \alpha <0.75$. 
The ground state has nontrivial fourfold degeneracy in the thermodynamic limit~\cite{Mandal2012}.

The exact solution for the ground state was originally obtained by introducing four types of Majorana fermions for each $S=1/2$ spin~\cite{Kitaev2006}. 
In this method, the Hilbert space in the original spin representation, $2^N$, is extended to $4^N$ in the Majorana fermion representation ($N$ is the number of spins).
Thus, to calculate physical quantities, such as spin correlations, it is necessary to make a projection from the extended Hilbert space to the original one. 

Soon later, however, another way of solving the model was introduced by using only two types of Majorana fermions~\cite{Chen2007,Feng2007,Chen2008}, in which the projection is avoided as the Hilbert space is not extended. 
In this method, the spin operators are written by spinless fermions by applying the Jordan-Wigner transformation to the one-dimensional chains composed of two types of bonds, say, the $J_x$ and $J_y$ bonds. Then, by introducing two Majorana fermions $c_j$ and $\bar{c}_j$ for the spinless fermions, the Hamiltonian in Eq.~(\ref{eq:Hamiltonian0}) is rewritten as
\begin{align}
\mathcal{H}=
i\frac{J_x}{4} \sum_{(j,j')_x} c_j c_{j'}
- i\frac{J_y}{4} \sum_{(j,j')_y} c_j c_{j'}
- i\frac{J_z}{4} \sum_{(j,j')_z} \eta_r c_j c_{j'},
\label{eq:Hamiltonian1}
\end{align}
where the sum over $(j,j')_p$ is taken for the NN sites on a $p$ bond with $j<j'$.
$\eta_r = i\bar{c}_j \bar{c}_{j'}$ is defined on each $z$ bond connecting $j$ and $j'$ sites ($r$ is the index of the $z$ bond).
Here, $\eta_r$ is considered as a $Z_2$ variable taking $\pm1$,  as $\eta_r$ commutes with the total Hamiltonian as well as with other $\eta_{r^\prime}$ and as $\eta_r^2 = 1$.
Thus, the model in Eq.~(\ref{eq:Hamiltonian1}) describes itinerant Majorana fermions $\{c_j\}$ (called matter fermions) coupled to the $Z_2$ variables $\{\eta_r\}$ (called gauge fluxes).
The ground state is given by all $\eta_r=1$, giving QSLs with gapless or gapful excitations depending on $\alpha$, as in the original Kitaev's solution.

In the present numerical study at finite $T$, we adopt the Majorana representation used in Eq.~(\ref{eq:Hamiltonian1}).
This is because the form of Eq.~(\ref{eq:Hamiltonian1}) is suitable for the CDMFT calculations (see Sec.~\ref{subsec:CDMFT}), as the interaction term, the third term in Eq.~(\ref{eq:Hamiltonian1}), only lies on $z$ bonds.
In this study, we apply the CDMFT to deal with thermal fluctuations and compute the static quantities. 
For calculating dynamical quantities, we apply the CTQMC method to the converged solutions obtained the CDMFT.
While the framework was briefly introduced in Ref.~\cite{Yoshitake2016}, we describe further details in the following sections.

\subsection{Cluster dynamical mean-field theory in the Majorana fermion representation\label{subsec:CDMFT}}
As presented in the previous study by the real-space QMC simulation~\cite{Nasu2015}, spacial correlations between $\eta_r$ develop at low $T$. 
To take into account such spacial correlations, we adopt a cluster extension of DMFT~\cite{Kotliar2001}.
As the Majorana Hamiltonian in Eq.~(\ref{eq:Hamiltonian1}) is formally similar to the Falicov-Kimball model or the double-exchange model with Ising localized moments, we follow the DMFT framework for the double-exchange model~\cite{Furukawa1994}.

In the CDMFT, we regard the whole lattice as a periodic array of clusters.
The Hamiltonian in Eq.~(\ref{eq:Hamiltonian1}) is rewritten into the matrix form of
\begin{align}
\mathcal{H} = 
\sum_{\gamma,\gamma^\prime,j,j^\prime} \frac{1}{2}\mathcal{H}^{0}_{\gamma,j;\gamma^\prime,j^\prime}c_{\gamma,j} c_{\gamma^\prime,j^\prime}
+ \sum_{\gamma,j,j^\prime} \frac{1}{2}\mathcal{H}^{\{\eta\}}_{j,j^\prime}c_{\gamma,j} c_{\gamma,j^\prime},
\label{eq:Hamiltonian2}
\end{align}
where $\gamma$ and $\gamma^\prime$ are the indices for the clusters, and $j$ and $j'$ denote the sites in each $N_c$-site cluster.
The coefficient $1/2$ in Eq.~(\ref{eq:Hamiltonian2}) is introduced to follow the notation in Ref.~\cite{Nilsson2013}.
In Eq.~(\ref{eq:Hamiltonian2}), the first term corresponds to the first and second terms in Eq.~(\ref{eq:Hamiltonian1}), while the second term is for the third term.
Green's function for Eq.~(\ref{eq:Hamiltonian2}) is formally written as 
\begin{align}
G(\mathbf{k}, i\omega_n)=(i\omega_n-2\mathcal{H}^{0}(\mathbf{k}) - \Sigma(\mathbf{k}, i\omega_n) )^{-1},
\label{eq:GreenFunc1}
\end{align}
where $\omega_n = (2n+1)\pi T$ is the Matsubara frequency ($n$ is an integer, and the Boltzmann constant $k_{\rm B}$ and the reduced Planck constant $\hbar$ are set to unity), $\Sigma(\mathbf{k}, i\omega_n)$ is the self-energy, and $\mathcal{H}^0(\mathbf{k})$ is the Fourier transform of $\mathcal{H}^{0}_{\gamma,j;\gamma^\prime,j^\prime}$ in Eq.~(\ref{eq:Hamiltonian2}) given by the matrix:
\begin{align}
\mathcal{H}^{0}_{j,j^\prime}(\mathbf{k}) = 
\sum_{\gamma} \mathcal{H}^{0}_{\gamma,j;0,j^\prime} e^{-i\mathbf{k}\cdot\mathbf{r}_{\gamma}},
\end{align}
where $\mathbf{r}_{\gamma}$ is the coordinate of the cluster $\gamma$.

Following the spirit of the DMFT~\cite{Metzner1989,Georges1996}, we omit the $\mathbf{k}$ dependence of the self-energy:
$\Sigma(\mathbf{k}, i\omega_n) = \Sigma(i\omega_n)$.
In this approximation, local Green's function is defined within a cluster as
\begin{align}
G_{j,j^\prime}(i\omega_n)= \frac{1}{N'}\sum_\mathbf{k} \left[(i\omega_n-2\mathcal{H}^{0}(\mathbf{k}) - \Sigma(i\omega_n) )^{-1}\right]_{j,j^\prime},
\label{eq:GreenFunc2}
\end{align}
where $N'$ is the number of clusters in the whole lattice ($N=N_c N'$),  and $j$ and $j'$ denotes the sites in the cluster.
The Weiss function is introduced to take into account the correlation effects in other clusters as
\begin{align}
\mathcal{G}^{0}_{j,j^\prime}(i\omega_n)^{-1} =
G_{j,j^\prime}(i\omega_n)^{-1}+ \Sigma_{j,j^\prime}(i\omega_n).
\label{eq:GreenFunc3}
\end{align}

In order to take into account the interaction $\mathcal{H}^{\{\eta\}}$ in Eq.~(\ref{eq:Hamiltonian2}) within the cluster that we focus on, we consider the impurity problem for the cluster described by the effective action in the path-integral representation for Majorana fermions~\cite{Nilsson2013}.
The partition function is given by
\begin{align} 
Z =\sum_{ {\{\eta\}} } Z^{\{\eta\}}, \label{eq:EffAction1}
\end{align}
where
\begin{align}
Z^{\{\eta\}} = \int \mathcal{D}\chi {\rm{exp}}(-\mathcal{S}_{\text{eff}}^{\{\eta\}}).
\label{eq:EffAction2}
\end{align}
Here, the sum of $\{\eta\}$ in Eq.~(\ref{eq:EffAction1}) runs over all possible configurations of $\{\eta\}$, and $\mathcal{D}\chi = \prod_{j,n} d\chi_{j,\omega_n}$ in Eq.~(\ref{eq:EffAction2}); 
$\chi_{j,\omega_n}$ is the Grassmann number corresponding to the Majorana operator $c_j$ (more precisely, $c_j/\sqrt{2}$ following the notation in Ref.~\cite{Nilsson2013}). 
The effective action is given by
\begin{align}
\mathcal{S}_{\text{eff}}^{\{\eta\}} =
& - T{\sum_{j,j',n \geq 0
}} \chi_{j,-\omega_n} (\mathcal{G}^{0}(i\omega_n))_{j,j'}^{-1} \chi_{j',\omega_n} \nonumber \\
&+2T{\sum_{j,j',n \geq 0}}\chi_{j,-\omega_n} \mathcal{H}_{j,j'}^{\{\eta\}} \chi_{j',\omega_n}.
\label{eq:EffAction3}
\end{align}
For a given configuration of $\{\eta\}$, the impurity problem defined by Eq.~(\ref{eq:EffAction2}) is exactly solvable because it is nothing but a free fermion problem. Green's function is obtained as
\begin{align}
\left[(G^{\{\eta\}}(i\omega_n))^{-1}\right]_{j,j'} = \left[(\mathcal{G}^{0}(i\omega_n))^{-1}\right]_{j,j'} - 2\mathcal{H}^{\{\eta\}}_{j,j'}.
\end{align} 
Note that we slightly modified the notation from the previous study in Ref.~\cite{Yoshitake2016}.
Then, local Green's function for the impurity problem is calculated by
\begin{align} 
G_{j,j'}^{\rm{imp}}(i\omega_n) = \sum_{ {\{\eta\}} } P(\{\eta\})G_{j,j'}^{\{\eta\}}(i\omega_n),
\label{eq:localG}
\end{align}
where $P(\{\eta\})$ is the statistical weight for the configuration $\{\eta\}$ given by
\begin{align}
P(\{\eta\}) = Z^{\{\eta\}}/\sum_{\{\eta\}}Z^{\{\eta\}}.
\end{align}
$Z^{\{\eta\}}$ is obtained from Green's functions as
\begin{align}
Z^{\{\eta\}}
=\prod_{n \geq 0} \text{det}[-G^{\{\eta\}}(i\omega_n)].
\end{align}
We note that $G^{\rm{imp}}(i\omega_n)$ is obtained exactly by computing $G^{\{\eta\}}(i\omega_n)$ and $P(\{\eta\})$ for all $2^{N_c/2}$ configurations of $\{\eta\}$ in the $N_c$-site cluster~\cite{Udagawa2012}.
The self-energy for the impurity problem is obtained as
\begin{align} 
\Sigma_{j,j'}(i\omega_n) = \left[(\mathcal{G}^{0}(i\omega_n))^{-1}\right]_{j,j'} - \left[(G^{\rm{imp}}(i\omega_n))^{-1}\right]_{j,j'}.
\label{eq:self-energy}
\end{align}

In the CDMFT, the above equations, Eqs.~(\ref{eq:GreenFunc2}), (\ref{eq:GreenFunc3}), (\ref{eq:localG}), and (\ref{eq:self-energy}), are solved in a self-consistent way.
The self-consistent condition is given by
\begin{align}
G(i\omega_n) = G^{\rm{imp}}(i\omega_n),
\end{align}
namely, the calculation is repeated until local Green's function in Eq.~(\ref{eq:GreenFunc2}) 
agrees with Green's function calculated for the impurity problem in Eq.~(\ref{eq:localG}).

The Majorana CDMFT framework provides a concise calculation method for $T$ dependences of  static quantities of the Kitaev model, such as the specific heat and the equal-time spin correlations $\langle S_j^p S_{j'}^p \rangle$. 
It is worth noting that the CDMFT calculations can be performed without any biased approximation except for the cluster approximation: the exact enumeration for all the $2^{N_c/2}$ configurations in Eq.~(\ref{eq:localG}) enables the exact calculations for the given cluster. 
Furthermore, the cluster-size dependence is sufficiently small at all the $T$ range above the critical temperature for the artificial phase transition due to the mean-field nature of the CDMFT, as demonstrated for the isotropic case with $\alpha=1.0$ in the previous study~\cite{Yoshitake2016} (see also Sec.~\ref{subsec:StaticQuantity} and Appendix~\ref{app:SizeDep} for the anisotropic cases). 
On the other hand, for obtaining dynamical quantities, such as the dynamical spin correlations $\langle S^{p}_{j}(\tau)S^{p}_{j'} \rangle$ ($\tau$ is the imaginary time), we need to make an additional effort beyond the exact enumeration in the CDMFT, as discussed in the next subsection.

In the CDMFT+CTQMC calculations in Sec.~\ref{sec:Result}, we use the 26-site cluster shown in Fig.~\ref{fig:fig1}(a).
In Appendix~\ref{app:SizeDep}, we examine the dependence on the cluster size as well as shape.

\subsection{Continuous-time quantum Monte Carlo method \label{subsec:CTQMC}}

In order to calculate the dynamical spin correlations $\langle S^{p}_{j}(\tau)S^{p}_{j'} \rangle$, we need to take into account the imaginary-time evolution of $\{\bar{c}\}$ that compose the conserved quantities $\{\eta\}$, e.g., $S^z_j(\tau) = \pm i \chi_j(\tau)\bar{c}_j(\tau)/\sqrt{2}$; the sign depends on the sublattice on the honeycomb structure.
For this purpose, we adopt the CTQMC method based on the strong coupling expansion~\cite{Werner2006}.
In this method, $\langle S^{z}_{j}(\tau)S^{z}_{j'} \rangle$ on an $r_0$ bond is calculated as
\begin{align}
\langle S^{z}_{j}(\tau)S^{z}_{j'} \rangle = \sum_{\{\eta\}', \eta_{r_0} = \pm 1}
P(\{\eta\}', \eta_{r_0}) \langle S^{z}_{j}(\tau)S^{z}_{j'} \rangle^{\{\eta\}' }, 
\label{eq:SztauSz}
\end{align}
where $\{\eta\}'$ represents the configurations of $\eta_r$ except for $\eta_{r_0}$ on the $r_0$ bond.
$P(\{\eta\}', \eta_{r_0})$ is obtained from the converged solution of the Majorana CDMFT in Sec.~\ref{subsec:CDMFT}. 
$\langle S^{z}_{j}(\tau)S^{z}_{j'} \rangle^{\{\eta\}' }$ is the dynamical spin correlation on the $r_0$ bond calculated by the CTQMC method for each configuration $\{\eta\}'$. 
The sum of $\{\eta\}'$ runs over all possible configurations of $\{\eta\}'$ within the cluster.  
Note that Eq.~(\ref{eq:SztauSz}) is derived from the fact that $S^{z}_{j}$ commutes with $\eta_r$ in $\{\eta\}'$, whereas it does not commute with $\eta_{r_0}$.
Thus, for a given $\{\eta\}'$, the interaction lies only on the $r_0$ bond, and hence, it is sufficient to solve the two-site impurity problem in the CTQMC calculations. 
The two-site impurity problem is defined by the integration in Eq.~(\ref{eq:EffAction2}) on $\chi_{j,\omega_n}$ whose $j$ does not belong to the $r_0$ bond.
Then, we obtain
\begin{align} 
\mathcal{S}_{\text{eff}}^{\{\eta\}'} =
\mathcal{S}_{\rm{hyb}}^{\{\eta\}'} + \mathcal{S}_{\rm{local}},
\label{eq:S_eff}
\end{align}
where
\begin{align}
\mathcal{S}_{\rm{hyb}}^{\{\eta\}'} =&-\sum_{j,j'} \int_0^\beta d\tau \int_0^\beta d\tau^{'} \chi_{j}(\tau)\Delta_{j,j'}^{\{\eta\}'}(\tau-\tau^{'})\chi_{j'}(\tau^{'}), \label{eq:ImTimeActionHyb}\\
\mathcal{S}_{\rm{local}} =& \sum_{j,j'} \int_0^\beta d\tau \chi_{j}(\tau) \left(\frac{\delta_{j,j'}}{2}\frac{\partial}{\partial\tau} + \mathcal{H}_{j,j'}^{\{\eta\}} \right) \chi_{j'}(\tau),
\label{eq:ImTimeActionLoc}
\end{align}
and $j, j'$ in Eqs.~(\ref{eq:ImTimeActionHyb}) and (\ref{eq:ImTimeActionLoc}) are the sites on the $r_0$ bond; $\beta = 1 / T$ is the inverse temperature. 
In Eq.~(\ref{eq:ImTimeActionHyb}), the hybridization function $\Delta_{j,j'}^{\{\eta\}'}(\tau)$ is calculated from $G_{j,j'}^{\{\eta\}}(i\omega_n)$ in the converged solution of CDMFT as follows.
Let us define the matrix $\tilde{G}^{\{\eta\}}(i\omega_n)$ as a $2\times 2$ submatrix of $G^{\{\eta\}}(i\omega_n)$, as
\begin{align}
\tilde{G}_{j,j'}^{\{\eta\}}(i\omega_n) = G_{j,j'}^{\{\eta\}}(i\omega_n).
\end{align}
Then, the hybridization function is given as a function of the Matsubara frequency in the form
\begin{align}
\Delta_{j,j'}^{\{\eta\}'}(i\omega_n) = [\tilde{G}^{\{\eta\}}(i\omega_n)]^{-1}_{j,j'} - ( i\omega_n - 2\mathcal{H}^{\{\eta\}}_{j,j'} ).
\label{eq:Delta_iwn}
\end{align} 
Note that $\Delta_{j,j'}^{\{\eta\}'}(i\omega_n)$ does not depend on $\{\eta\}$, which is straightforwardly shown by the matrix operations in the right hand side.
Converting Eq.~(\ref{eq:Delta_iwn}) to the imaginary-time representation, we obtain
\begin{align}
\Delta_{j,j'}^{\{\eta\}'}(\tau) = \frac{T}{2}\sum_n e^{-i\omega_n \tau} \Delta_{j,j'}^{\{\eta\}'}(i\omega_n).
\end{align}
Given Eqs.~(\ref{eq:S_eff})-(\ref{eq:ImTimeActionLoc}), the partition function of the system is expanded in terms of $\mathcal{S}_{\rm{hyb}}^{\{\eta\}'}$ as
\begin{widetext}
\begin{align} 
\frac{Z}{Z_{\rm{local}}} &= \frac{\int \mathcal{D}\chi e^{-\mathcal{S}_{\rm{hyb}}^{\{\eta\}'}}e^{- \mathcal{S}_{\rm{local}}} }
{ \int \mathcal{D}\chi e^{ - \mathcal{S}_{\rm{local}}}}
= \langle e^{-\mathcal{S}_{\rm{hyb}}^{\{\eta\}'}} \rangle_{\rm{local}} \nonumber\\
&=\sum_{d, i_0,..., i_{2d-1}} \int_{0}^{\beta} d\tau_0 ... \int_{0}^{\beta} d\tau_{2d-1} 
\frac{1}{d!} \langle
\chi_{i_0}(\tau_0) ... \chi_{i_{2d-1}}(\tau_{2d-1}) \rangle_{\rm{local}}
{\rm Pf}(\hat{\Delta}^{\{\eta\}'}(d, i_0, \tau_0,..., i_{2d-1}, \tau_{2d-1})),
\label{eq:Z_CTQMC}
\end{align}
\end{widetext}
where $Z_{\rm{local}} = \int \mathcal{D}\chi e^{ - \mathcal{S}_{\rm{local}}}$ is the partition function for the two sites described by $\mathcal{S}_{\rm{local}}$, and $\langle\mathcal{A}\rangle_{\rm{local}}$ represents the expectation value in the two-site problem as
\begin{align}
\langle\mathcal{A}\rangle_{\rm{local}} = \frac{\int \mathcal{D}\chi
\mathcal{A}e^{ - \mathcal{S}_{\rm{local}}}}{\int \mathcal{D}\chi e^{ - \mathcal{S}_{\rm{local}}}}.
\end{align}
In the second line of Eq.~(\ref{eq:Z_CTQMC}), $d$ is the order of $\mathcal{S}_{\rm{hyb}}^{\{\eta\}'}$ in the expansion of $e^{-\mathcal{S}_{\rm{hyb}}^{\{\eta\}'}}$, Pf($M$) is the Pfaffian of skew-symmetric matrix $M$, and $\hat{\Delta}^{\{\eta\}'}(d, i_0, \tau_0,..., i_{2d-1}, \tau_{2d-1})$ is a $2d \times 2d$ matrix, whose $(m,n)$ element is given by 
\begin{align}
\hat{\Delta}^{\{\eta\}'}(d, i_0, \tau_0..., i_{2d-1}, \tau_{2d-1})_{m,n} = \Delta_{i_m, i_n}^{\{\eta\}'}(\tau_m-\tau_n).
\end{align}
We note that this is the first formulation of the CTQMC method with using the Pfaffian in the weight function to our knowledge, whereas a QMC simulation in the Majorana representation has been introduced for itinerant fermion models~\cite{Li2015}.

In the CTQMC calculation, we perform MC sampling over the configurations $(d, i_0,\tau_0,..., i_{2d-1},\tau_{2d-1})$ by using the integrand in Eq.~(\ref{eq:Z_CTQMC}) as the statistical weight for each configuration.
In each MC step, we perform an update from one configuration to another; for instance, an increase of the order of expansion $d$ as $(d, i_0,\tau_0,..., i_{2d-1},\tau_{2d-1})$ to $(d+1, i_0,\tau_0,..., i_{2d-1},\tau_{2d-1}, i_{2d}, \tau_{2d}, i_{2d+1},\tau_{2d+1})$ by adding $(i_{2d},\tau_{2d}),(i_{2d+1},\tau_{2d+1})$.
To judge the acceptance of such an update, we need to calculate the ratio of the Pfaffian.
This is efficiently done by using the fast update algorithm, as in the hybridization expansion scheme for usual fermion problems (for example, see Ref.~\cite{Rubtsov2005}).
For the above example of increasing $d$, the ratio is calculated by adding two rows and columns in the matrix $\hat{\Delta}^{\{\eta \}'}$ as
\begin{align} 
\frac{{\rm Pf}(\hat{\Delta}^{\{\eta\}'}(d, i_0, \tau_0,..., i_{2d-1}, \tau_{2d-1}))}
{{\rm Pf}(\hat{\Delta}^{\{\eta\}'}(d+1, i_0, \tau_0,..., i_{2d+1}, \tau_{2d+1}))},
\end{align}
whose calculation cost is in the order of $d^2$ by using the fast update algorithm.
On the other hand, in Eq.~(\ref{eq:Z_CTQMC}),
$\langle {T_\tau \chi_{i_0}(\tau_0) ... \chi_{i_{2d-1}}(\tau_{2d-1}) } \rangle_{\rm{local}}$ is obtained as the average in the two-site problem, which can be calculated by considering the imaginary-time evolution of all the four states in the two-site problem.

Then, the dynamical spin correlation for the configuration $\{\eta\}'$, $\langle S^z_j(\tau)S^z_{j'} \rangle^{\{\eta\}'}$ in Eq.~(\ref{eq:SztauSz}), is calculated as
\begin{widetext}
\begin{align} 
\langle S^z_j(\tau)S^z_{j'} \rangle^{\{\eta\}'} 
=\frac{Z_{\rm{local}}}{Z}\sum_{d, i_0,...,i_{2d-1}} \int_{0}^{\beta} d\tau_0 ... \int_{0}^{\beta} d\tau_{2d-1}
\frac{1}{d!} &\langle {\chi_{i_0}(\tau_0) ... \chi_{i_{2d-1}}(\tau_{2d-1}) S_j^{z}(\tau)S_{j'}^{z}} \rangle_{\rm{local}}
\nonumber \\
& \times {\rm Pf}(\hat{\Delta}^{\{\eta\}'}(d, i_0,..., i_{2d-1}, \tau_0, ..., \tau_{2d-1})).
\end{align}
\end{widetext}
For the MC sampling, we need to evaluate 
\begin{align} 
\frac{\langle {T_\tau \chi_{i_0}(\tau_0) ... \chi_{i_{2d-1}}(\tau_{2d-1}) S^z_j(\tau)S^z_{j'}} \rangle_{\rm{local}} }{\langle {T_\tau \chi_{i_0}(\tau_0) ... \chi_{i_{2d-1}}(\tau_{2d-1})} \rangle_{\rm{local}}}.
\end{align}
This is again calculated by considering the imaginary-time evolution of all the four states in the two-site problem.
In the isotropic case with $\alpha=1.0$, $\langle S^p_j(\tau)S^p_{j'} \rangle$ for $p=x,y$ are equivalent to $\langle S^z_j(\tau)S^z_{j'} \rangle$. 
Meanwhile, for the anisotropic case, we compute $\langle S^p_j(\tau)S^p_{j'} \rangle$ for $p=x,y$ by the same technique described above with using the spin rotations $\{S^x,S^y,S^z\} \to \{S^y, S^z, S^x \}$ or $\{S^x,S^y,S^z\} \to \{S^z, S^x,S^y\}$. 

In the CTQMC calculations in Sec.~\ref{sec:Result}, for each configuration $\{\eta\}'$, we typically run $10^7$ MC steps and perform the measurements at every 20 steps, after $10^5$ MC steps for the initial relaxation.

\subsection{Maximum entropy method\label{subsec:MEM}}

By using the CTQMC method as the impurity solver in the CDMFT, which we call the CDMFT+CTQMC method, we can numerically estimate the dynamical spin correlation as a function of the imaginary time, $\langle S^{p}_{j}(\tau)S^{p}_{j'} \rangle$.
To obtain the physical observables, such as the dynamical spin structure factor and the NMR relaxation rate, which are given by the dynamical spin correlations as functions of frequency $\omega$, we need to inversely solve the equation given by the generic form $g(\tau) = \int d\omega \rho(\omega)e^{-\omega\tau}$. 
In our problem, $g(\tau)$ and $\rho(\omega)$ correspond to the dynamical spin correlations as functions of imaginary time $\tau$ and real frequency $\omega$: 
$g(\tau) = \langle S^{p}_{j}(\tau)S^{p}_{j'} \rangle$ and $\rho(\omega) =  S^{p}_{j,j'}(\omega)$.
In the following calculations, we utilize the Legendre polynomial expansion following Refs.~\cite{Boehnke2011,Levy2017}: 
\begin{align}
g_{m}=\sqrt{2m+1}\int_0^\beta d\tau P_{m}(x(\tau))g(\tau),
\end{align}
where $P_m(x)$ is the $m$th Legendre polynomials and $x(\tau) = 2\tau/\beta - 1$. 
Then, the inverse problem is given by
\begin{align}
g_m = \int d\omega \rho(\omega)K_{m}(\omega),
\end{align}
where
\begin{align}
K_{m}(\omega)=\sqrt{2m+1}\int_0^\beta d\tau P_{m}(x(\tau))e^{-\omega\tau}.
\end{align}

For solving the inverse problem, we adopt the maximum entropy method (MEM)~\cite{Jarrell1996}.
The following procedure is the standard one, but we briefly introduce it to make the paper self-contained. 
In the MEM, we discretize $\rho(\omega)$ to $\rho_l=\rho(\omega_l)$, and determine $\rho_l$ to minimize the function
\begin{align} 
\mathcal{F} &= \frac{1}{2}\sum_{m,n} (g_m - \tilde{g}_m)\zeta^{-1} C^{-1}_{m,n}(g_n - \tilde{g}_n) \nonumber \\
&- \delta\sum_l\left[\rho_l - \rho^{(0)}_l - \rho_l {\rm ln}\left(\frac{\rho_l}{\rho^{(0)}_l}\right) \right], \label{eq:MaxEnt}
\end{align}
where $\zeta$ and $\delta$ are the coefficients described below, and $C$ is a variance-covariance matrix of $g_m$; $\tilde{g}_m = \sum_l \Delta\omega \rho_l K_{m}(\omega_l)$. 
We take the Legendre expansion up to $50$th order and $\Delta \omega = 0.01125$ in the following calculations. 
In Eq.~(\ref{eq:MaxEnt}), $\rho^{(0)}_l$ is the advance estimate of $\rho_l$, which we set to be a constant in this study.

Once neglecting the second term in the right hand side of Eq.~(\ref{eq:MaxEnt}), the minimization of $\mathcal{F}$ is equivalent to the least squares method.
The least squares method is unstable, as $g_m$ is rather insensitive to a change of $\rho_l$.
The second term, called the entropy term, stabilizes the minimization process. 
In the following calculations, we set $\zeta=625$ to sufficiently take into account the effect of the entropy term, where the value of $\delta$ is determined self-consistently in each MEM calculation based on the maximum likelihood estimation, called the classical MEM~\cite{Jarrell1996} (typically, $\delta\simeq 1$-$10$). 
We note that the deviations of $\tilde{g}_m$ from $g_m$ are typically
comparable to the statistical errors in the CTQMC calculations.
In the following results, we estimate the errors of $\rho(\omega)$ by the standard deviation between the data for $\zeta=100$, $625$ and $10000$ in the range where the MEM retains the precision. 

In the MEM, $\rho(\omega)$ should be positive for all $\omega$. 
In our problem, the onsite correlation $S^{p}_{j,j}(\omega)$ satisfies this condition automatically, whereas $S^{p}_{j,j'}(\omega)$ for the NN sites $j,j'$ on the $p$ bond, which is denoted by $S^{p}_{\rm{NN}}(\omega)$ hereafter, can be negative.
(Note that all the further-neighbor correlations beyond the NN sites vanish in the Kitaev model~\cite{Baskaran2007}.)
To obtain $S^{p}_{\rm{NN}}(\omega)$ properly, we calculate $S^{p}_{j,j}(\omega) + 2S^{p}_{j,j'}(\omega) + S^{p}_{j',j'}(\omega)$, which is positive definite for all $\omega$, and subtract the onsite contributions~\cite{note}.
The accuracy of $S^{p}_{j,j'}(\omega)$ obtained by the MEM are examined in Appendix~\ref{app:MEM} in the one-dimensional limit with $\alpha = 1.5$, where $S^{p}_{j,j'}(\omega)$ can be calculated without using the MEM.

\section{Result \label{sec:Result}}

In this section, we present the results obtained by the CDMFT and the CDMFT+CTQMC methods.
In Sec.~\ref{subsec:StaticQuantity}, we present the specific heat and equal-time spin correlations for the NN sites obtained by the CDMFT for the cases with anisotropic $J_x$, $J_y$, and $J_z$.
By comparing the results with those by the QMC method~\cite{Nasu2015}, we confirm that the CDMFT is valid in the $T$ range above the artificial critical temperature close to the low-$T$ crossover.
In Sec.~\ref{subsec:DSF}, \ref{subsec:RelaxationRate}, and \ref{subsec:Susceptibility}, we present the CDMFT+CTQMC results for dynamical quantities, i.e., the dynamical spin structure factor, the NMR relaxation rate, and the magnetic susceptibility, respectively, in the qualified $T$ range.
We discuss the results in comparison with the isotropic case reported previously in Ref.~\cite{Yoshitake2016}.

\subsection{Static quantities: comparison to the previous QMC results\label{subsec:StaticQuantity}}

\begin{figure}[t]
    \includegraphics[width=0.8\columnwidth,clip]{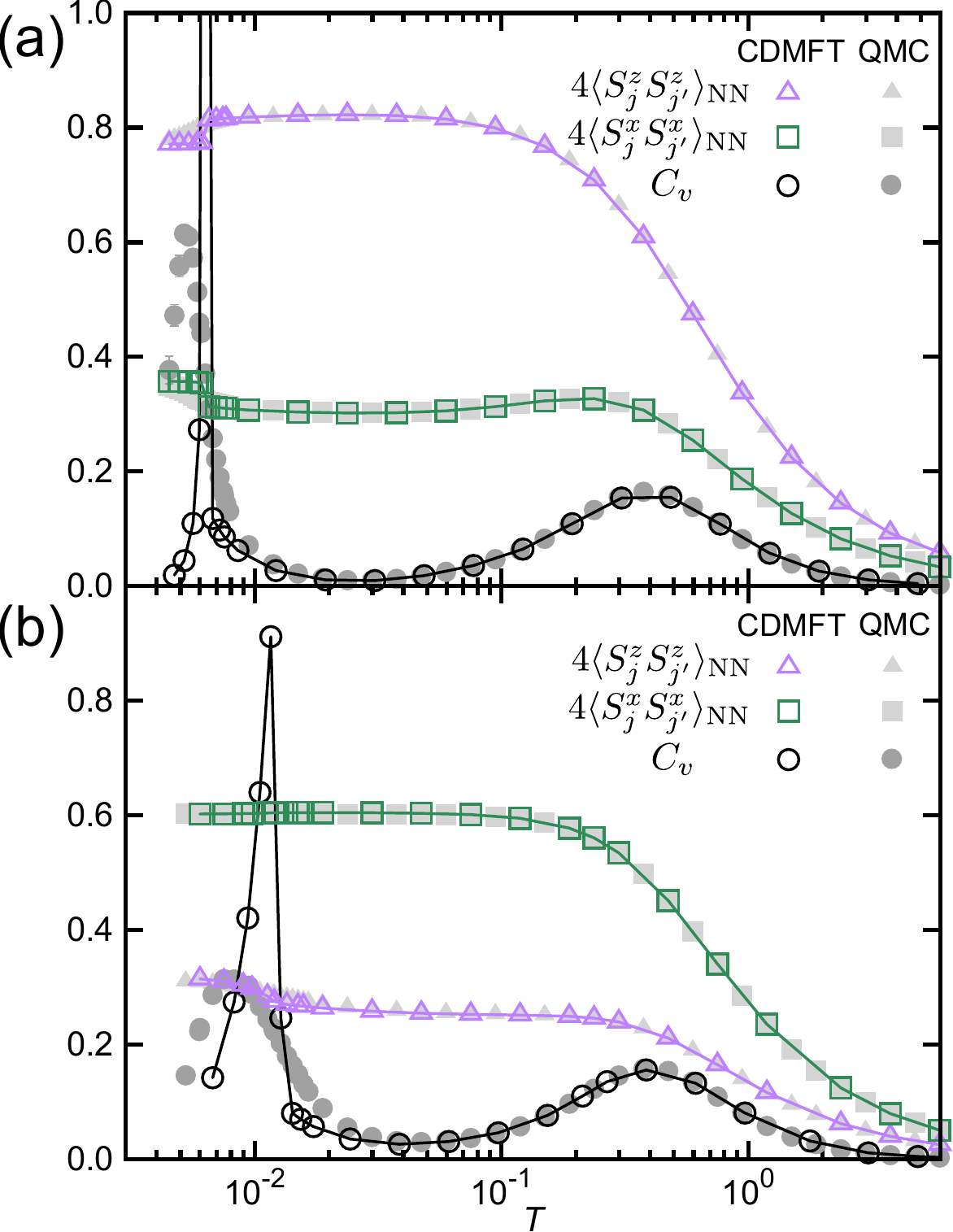}
    \caption{ \label{fig:fig2}
    The specific heat $C_v$ and equal-time spin correlations for the NN sites,
    $\langle S_j^z S_{j'}^z \rangle_{\rm{NN}}$ and $\langle S_j^x S_{j'}^x \rangle_{\rm{NN}}$, obtained by the Majorana CDMFT for the FM case at (a) $\alpha=0.8$ and (b) $\alpha=1.2$.
    Note that $\langle S_j^x S_{j'}^x \rangle_{\rm{NN}} = \langle S_j^y S_{j'}^y \rangle_{\rm{NN}}$ from the symmetry.
    QMC data in Ref.~\cite{Nasu2015} are plotted by gray symbols for comparison.
    }
\end{figure}

Figure~\ref{fig:fig2} shows the benchmark of the Majorana CDMFT.
We compare the specific heat $C_v$ and equal-time spin correlations for NN pairs on the $p$ bonds, $\langle S_j^p S_{j'}^p \rangle_{\rm{NN}}$, obtained by the Majorana CDMFT, with those by QMC in Ref.~\cite{Nasu2015}.
The data are calculated for the FM case with bond asymmetry: 
$\alpha=0.8$ ($J_x=J_y=0.8$ and $J_z=1.4$) and $\alpha=1.2$ ($J_x=J_y=1.2$ and $J_z=0.6$). 
While the data of $C_v$ are common to the FM and AFM cases, the sign of $\langle S_j^p S_{j'}^p \rangle_{\rm{NN}}$ is reversed for the AFM case.
Note that similar comparison was made for the isotropic case $\alpha=1.0$ ($J_x=J_y=J_z=1$) in Ref.~\cite{Yoshitake2016}.

As indicated by two broad peaks in the specific heat in the QMC results, the system exhibits two crossovers owing to thermal fractionalization of quantum spins~\cite{Nasu2015}; 
the crossover temperatures were estimated as $T_{\rm{L}} \simeq 0.012$ and $T_{\rm{H}} \simeq 0.375$ in the isotropic case.
In the anisotropic cases, the low-$T$ crossover takes place at a lower $T$, i.e., $T_{\rm{L}} \simeq 0.0052$ for $\alpha=0.8$ and $T_{\rm L} \simeq 0.0075$ for $\alpha=1.2$, while the high-$T$ one is almost unchanged, i.e., $T_{\rm{H}} \simeq 0.375$.
These behaviors are excellently reproduced by the Majorana CDMFT, except for the low-$T$ peak;
the CDMFT results show a sharp anomaly at $\tilde{T}_{\rm c} \simeq 0.0063$ for $\alpha=0.8$ and $\tilde{T}_{\rm c} \simeq 0.013$ for $\alpha=1.2$.
This is due to a phase transition by ordering of $\eta$, which is an artifact of the mean-field nature of CDMFT. 

On the other hand, the QMC results for the NN spin correlations are also precisely reproduced by the Majorana CDMFT in the wide $T$ range above the artificial phase transition temperature $\tilde{T}_{\rm c}$.
Although they appear to be reproduced even below $\tilde{T}_{\rm c}$, there is a small anomaly at $\tilde{T}_{\rm c}$ associated with the artificial transition, while the QMC data smoothly change around $T_{\rm L}$.
(Note that the appropriate sum of the NN spin correlations is nothing but the internal energy, and hence, the $T$ derivative corresponds to the specific heat.)

Thus, the comparison indicates that the Majorana CDMFT gives quantitatively precise results in the wide $T$ range above the artificial transition temperature $\tilde{T}_{\rm c}$: 
in the present cases with $\alpha=0.8$ and $1.2$, the CDMFT is reliable for $T \gtrsim 0.007$ and $T \gtrsim 0.014$, respectively.
As discussed in the previous study~\cite{Nasu2015}, the thermal fractionalization of quantum spins sets in below $T\simeq T_{\rm H}$, which is well above $\tilde{T}_{\rm c}$. 
Thus, the $T$ ranges qualified for the CDMFT include the peculiar paramagnetic state showing the thermal fractionalization.
In the following sections, we apply the CDMFT+CTQMC method in these qualified $T$ ranges to the study of spin dynamics, which was not obtained by the previous QMC method~\cite{Nasu2015}.

\subsection{Dynamical spin structure factor\label{subsec:DSF}}

\begin{figure*}[t]
    \includegraphics[width=2\columnwidth,clip]{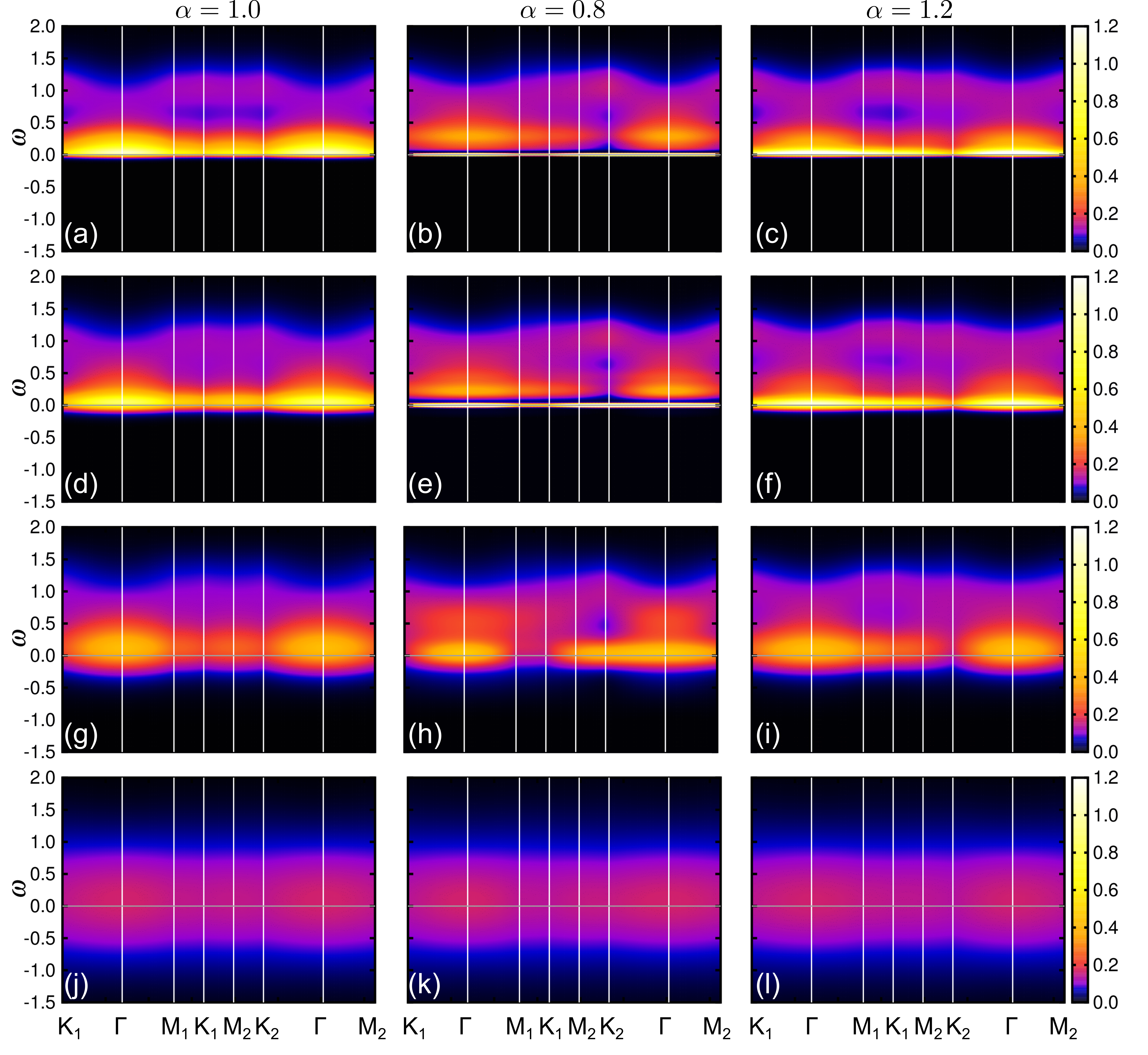}
    \caption{ \label{fig:fig3}
    Dynamical spin structure factor $S(\mathbf{q}, \omega)$ obtained by the Majorana CDMFT+CTQMC method for the FM case with (a)(d)(g)(j) $\alpha=1.0$, (b)(e)(h)(k) $\alpha=0.8$, and (c)(f)(i)(l) $\alpha=1.2$: (a)(b)(c) $T\simeq 2T_{\rm{L}}$, (d)(e)(f) $T\simeq \sqrt{T_{\rm{L}}T_{\rm{H}}}$, (g)(h)(i) $T\simeq 0.64T_{\rm{H}}$, and (j)(k)(l) $T\simeq 6.4T_{\rm{H}}$.
    Here, $T_{\rm L} \simeq 0.012$, $0.0052$, and $0.0075$ for $\alpha=1.0$, $0.8$, and $1.2$, respectively, while $T_{\rm H} \simeq 0.375$ for all the cases.
    }
\end{figure*}

Figure~\ref{fig:fig3} shows the CDMFT+CTQMC results for the dynamical spin structure factor $S(\mathbf{q}, \omega)$ at several $T$ for the FM case with $\alpha = 1.0$, $0.8$, and $1.2$. 
$S(\mathbf{q}, \omega)$ is calculated as
\begin{align}
S(\textbf{q}, \omega)&=\sum_p S^{p}(\textbf{q}, \omega), \\
S^{p}(\textbf{q}, \omega)&=\frac{1}{3N} \sum_{j,j'} e^{i\textbf{q}\cdot(\textbf{r}_{j}-\textbf{r}_{j'})}S^{p}_{j,j'}(\omega),
\label{eq:S^p(q,w)}
\end{align}
where $S^{p}_{j,j'}(\omega)$ is obtained by the MEM described in Sec.~\ref{subsec:MEM} from the imaginary-time correlations $\langle S^{p}_{j}(\tau)S^{p}_{j'} \rangle$ by CDMFT+CTQMC. 
As mentioned above, nonzero contributions in Eq.~(\ref{eq:S^p(q,w)}) come from only the onsite and NN-site components of $S^{p}_{j,j'}(\omega)$; 
we present their $T$ dependences in Appendix~\ref{app:S(w,T)}. 
The Brillouin zone and symmetric lines on which $S(\mathbf{q}, \omega)$ is plotted are presented in Fig.~\ref{fig:fig1}(b).
Although the results at $\alpha = 1.0$ were shown in the previous study~\cite{Yoshitake2016}, we present them (for a sightly different $T$ set) for comparison.
We show the data at four temperatures: $T\simeq 2T_{\rm L}$, $\sqrt{T_{\rm L} T_{\rm H}}$, $0.64 T_{\rm H}$, and $6.4 T_{\rm H}$. 
Note that $T_{\rm L} \simeq 0.012$, $0.0052$, and $0.0075$ for $\alpha=1.0$, $0.8$, and $1.2$, respectively, while $T_{\rm H} \simeq 0.375$ for all the cases.

As shown in Fig.~\ref{fig:fig3}, at sufficiently high $T$ than $T_{\rm{H}}$, $S(\mathbf{q}, \omega)$ does not show any significant $\mathbf q$ dependence for all $\alpha$ studied here; 
$S(\mathbf{q}, \omega)$ shows only a diffusive response centered at $\omega \sim 0$, as shown in Figs.~\ref{fig:fig3}(j)-\ref{fig:fig3}(l).
When lowering $T$ below $T_{\rm{H}}$, the diffusive weight is shifted to the positive $\omega$ region ranging up to above $\omega \sim J$ for all the cases, as shown in Figs.~\ref{fig:fig3}(g)-\ref{fig:fig3}(i).
Simultaneously, a quasi-elastic component grows gradually at $\omega \sim 0$.
Both the inelastic and the quasi-elastic components show a discernible $\mathbf{q}$ 
dependence; 
in particular, the latter increases the intensity around the $\Gamma$ point reflecting the FM interactions. 
While $S({\rm{K}}_1, \omega) = S({\rm{K}}_2, \omega)$ and $S({\rm{M}}_1, \omega) = S({\rm{M}}_2, \omega)$ for $\alpha=1.0$ from the symmetry, the quasi-elastic response is small (large) around the M$_1$-K$_1$ line compared to that around the M$_2$-K$_2$ line for $\alpha = 0.8$ ($1.2$) because of the anisotropy.

When further lowering $T$ and approaching $T_{\rm L}$, the quasi-elastic component increases its intensity, while the inelastic response at $\omega \sim J$ does not change substantially.
In particular, in the case of $\alpha = 0.8$, the quasi-elastic component is sharpened and develops to a $\delta$-function like peak as shown in Figs.~\ref{fig:fig3}(e) and \ref{fig:fig3}(b).
In addition, the broad incoherent weight splits from the coherent peak.
These behaviors appear to asymptotically converge onto the result at $T=0$, where the $\delta$-function peak appears due to the change of the parity between the ground state and the flux-excited state~\cite{Knolle2014} (for the $\delta$-function peak, see also Fig.~\ref{fig:fig16} in Appendix~\ref{app:MEM}). 
On the other hand, $S(\mathbf{q}, \omega)$ at $\alpha = 1.2$ does not show such a drastic change, and the quasi-elastic component grows continuously, as shown in Figs.~\ref{fig:fig3}(f) and \ref{fig:fig3}(c).
We note that the results for $\alpha=1.2$ are qualitatively similar to those for $\alpha=1.0$ in Figs.~\ref{fig:fig3}(d) and \ref{fig:fig3}(a), except for different $\mathbf{q}$ dependence mentioned above.

\begin{figure*}[t]
    \includegraphics[width=2\columnwidth,clip]{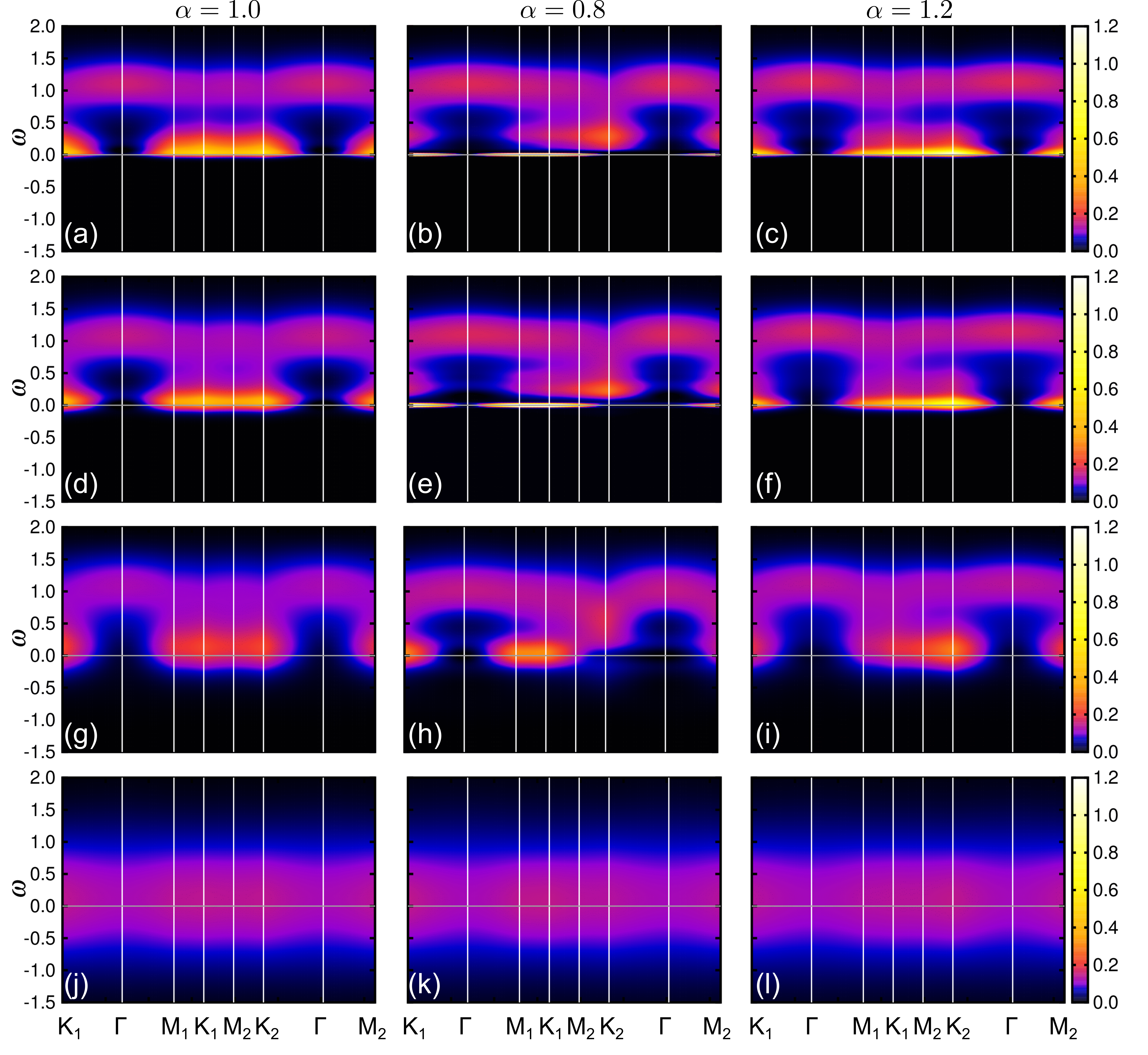}
    \caption{ \label{fig:fig4}
    Dynamical spin structure factor $S(\mathbf{q}, \omega)$ obtained by the Majorana CDMFT+CTQMC method for the AFM case.
    The values of $\alpha$ and $T$ are common to Fig.~\ref{fig:fig3}.
    }
\end{figure*}

Figure~\ref{fig:fig4} shows the results for the AFM case.
The overall $\omega$ dependence of $S(\mathbf{q}, \omega)$ is similar to that for the FM case at all $T$: 
the diffusive response centered at $\omega \sim 0$ for $T\gtrsim T_{\rm{H}}$ [Figs.~\ref{fig:fig4}(j)-\ref{fig:fig4}(l)], 
the shift of the diffusive weight to the region of $\omega \sim J$ and the growth of a quasi-elastic component at $\omega \sim 0$ below $T_{\rm{H}}$ [Figs.~\ref{fig:fig4}(g)-\ref{fig:fig4}(i)], and the $\delta$-function like peak for $\alpha = 0.8$ while approaching to $T_{\rm{L}}$ [Figs.~\ref{fig:fig4}(e)  and \ref{fig:fig4}(b)]. 
The similarity of the $\omega$ dependences of $S(\mathbf{q}, \omega)$ between FM and AFM cases is partly understood by the relation $2S({\rm{K}}_1,\omega)_{\rm{FM}} + S({\rm{K}}_{2},\omega)_{\rm{FM}} = 2S({\rm{K}}_1,\omega)_{\rm{AFM}} + S({\rm{K}}_{2},\omega)_{\rm{AFM}}$, which holds for $J_x = J_y$ [$S(\mathbf{q},\omega)_{\rm{FM}}$ and $S(\mathbf{q},\omega)_{\rm{AFM}}$ are $S(\mathbf{q},\omega)$ for the FM and AFM cases, respectively].
On the other hand, the $\mathbf{q}$ dependence is in contrast to the FM case: 
while the weight of the quasi-elastic response almost vanishes around the $\Gamma$ point, those on the zone boundary are enhanced in an almost opposite manner to the FM cases.
In addition, the incoherent weight at $\omega \sim J$ also shows the opposite $\mathbf{q}$ dependence to the FM case:
the weight is stronger around the $\Gamma$ point than that on the zone boundary.
The opposite $\mathbf{q}$ dependences between the FM and AFM cases directly follow
from the relation $S(\textbf{q},\omega)_{\rm{AFM}} = -S(\textbf{q},\omega)_{\rm{FM}} + (2/3)\sum_p S^p_{j,j}(\omega)$.

\begin{figure}[t]
    \includegraphics[width=\columnwidth,clip]{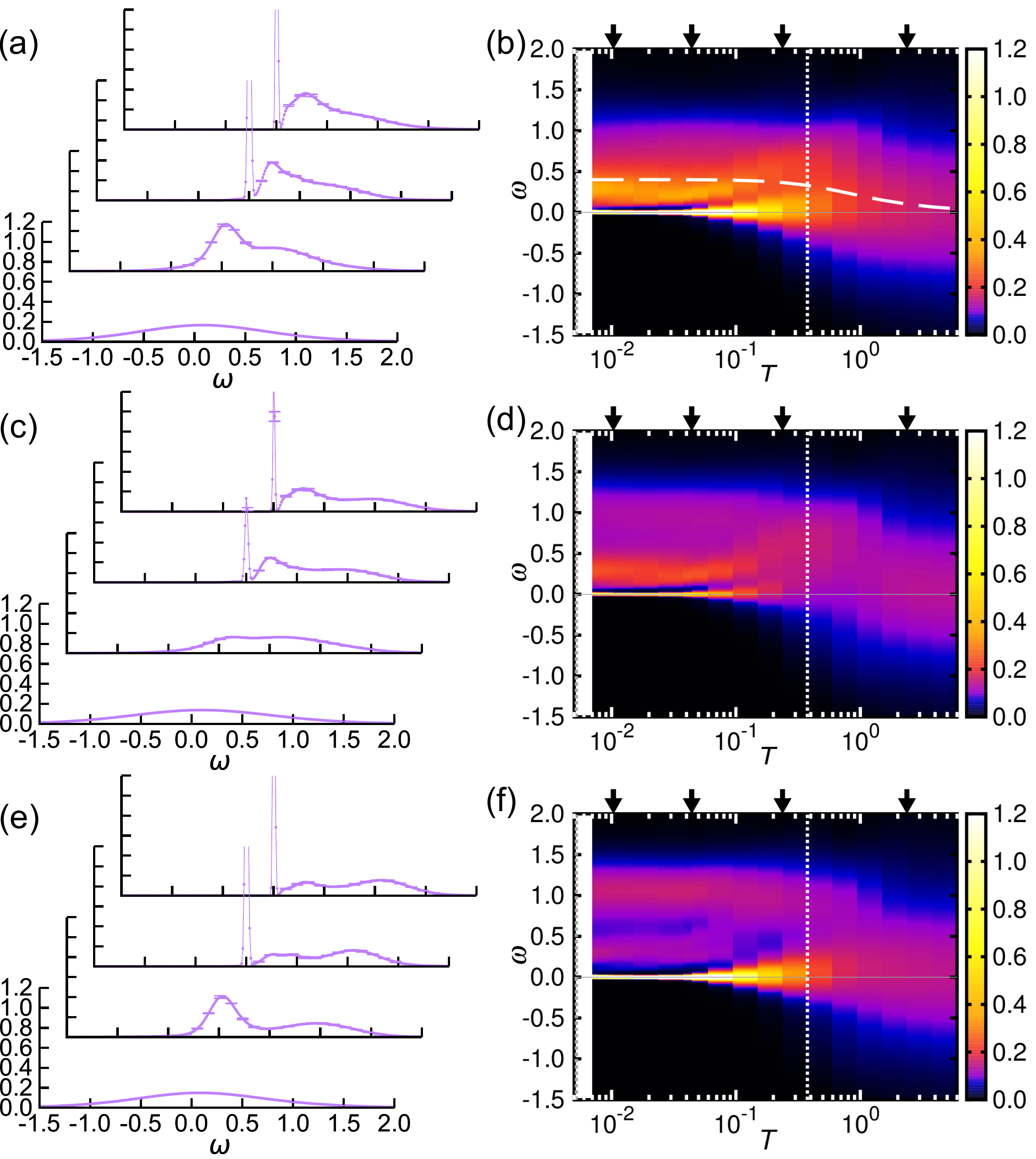}
    \caption{ \label{fig:fig5}
    (a) $S(\Gamma,\omega)$, (c) $S({\rm{K}_1},\omega)$, and (e) $S({\rm{K}_2},\omega)$ for the  FM case with $\alpha=0.8$
    at several $T$.
    The corresponding contour plots in the $T$-$\omega$ plane are shown in (b)(d)(f).
    The arrows indicate the temperatures used for the data in (a)(c)(e), while the white and gray dotted lines indicate $T_{\rm H}$ and $T_{\rm L}$, respectively.
    Note that the $T$ set is common to that used in Figs.~\ref{fig:fig3} and \ref{fig:fig4}. 
    The dashed curve in (b) represent the average frequency of  $S(\Gamma, \omega)$ (see the text for details).
    In (a)(c)(e), the errorbars are shown for every ten data along the $\omega$ axis. 
    }
\end{figure}

\begin{figure}[t]
    \includegraphics[width=\columnwidth,clip]{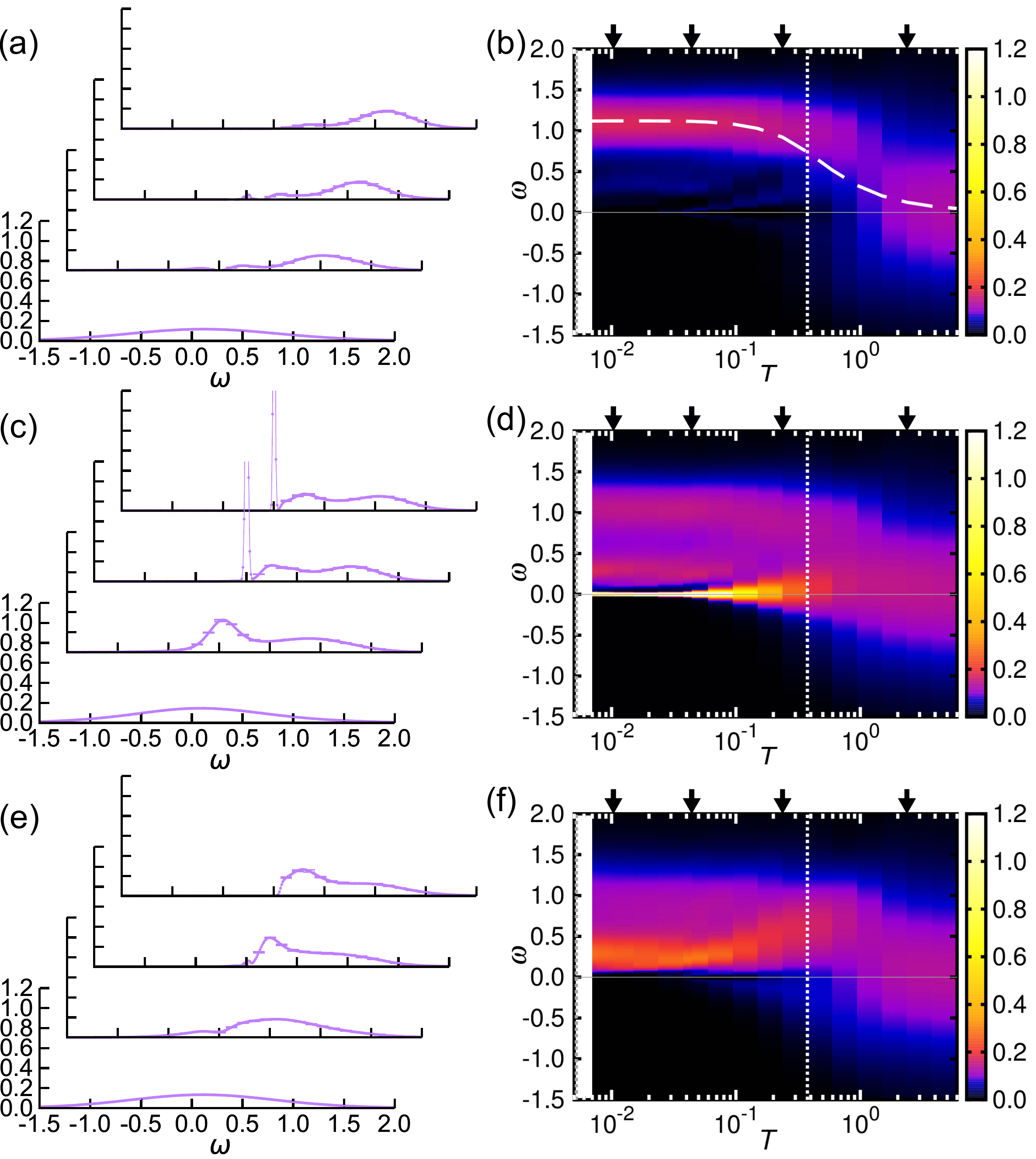}
    \caption{ \label{fig:fig6}
        (a) $S(\Gamma,\omega)$, (c) $S({\rm{K}_1},\omega)$, and (e) $S({\rm{K}_2},\omega)$ for the  AFM case with $\alpha=0.8$ at several $T$.
    The corresponding contour plots in the $T$-$\omega$ plane are shown in (b)(d)(f).
    The notations are common to those in Fig.~\ref{fig:fig5}.
    }
\end{figure}

\begin{figure}[t]
    \includegraphics[width=\columnwidth,clip]{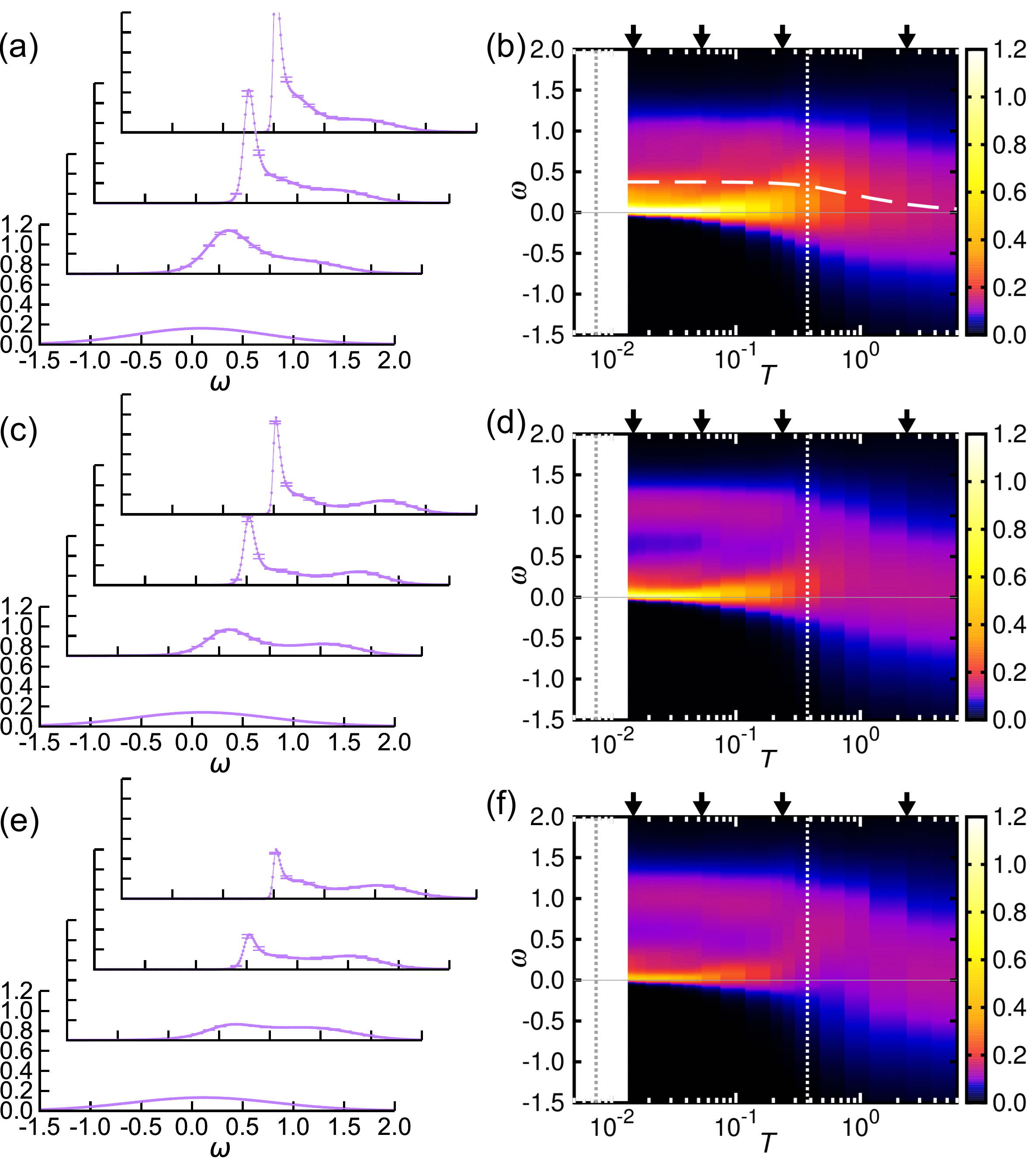}
    \caption{ \label{fig:fig7}
        (a) $S(\Gamma,\omega)$, (c) $S({\rm{K}_1},\omega)$, and (e) $S({\rm{K}_2},\omega)$ for the  FM case with $\alpha=1.2$ at several $T$.
    The corresponding contour plots in the $T$-$\omega$ plane are shown in (b)(d)(f).
    The notations are common to those in Fig.~\ref{fig:fig5}.
    }
\end{figure}

\begin{figure}[t]
    \includegraphics[width=\columnwidth,clip]{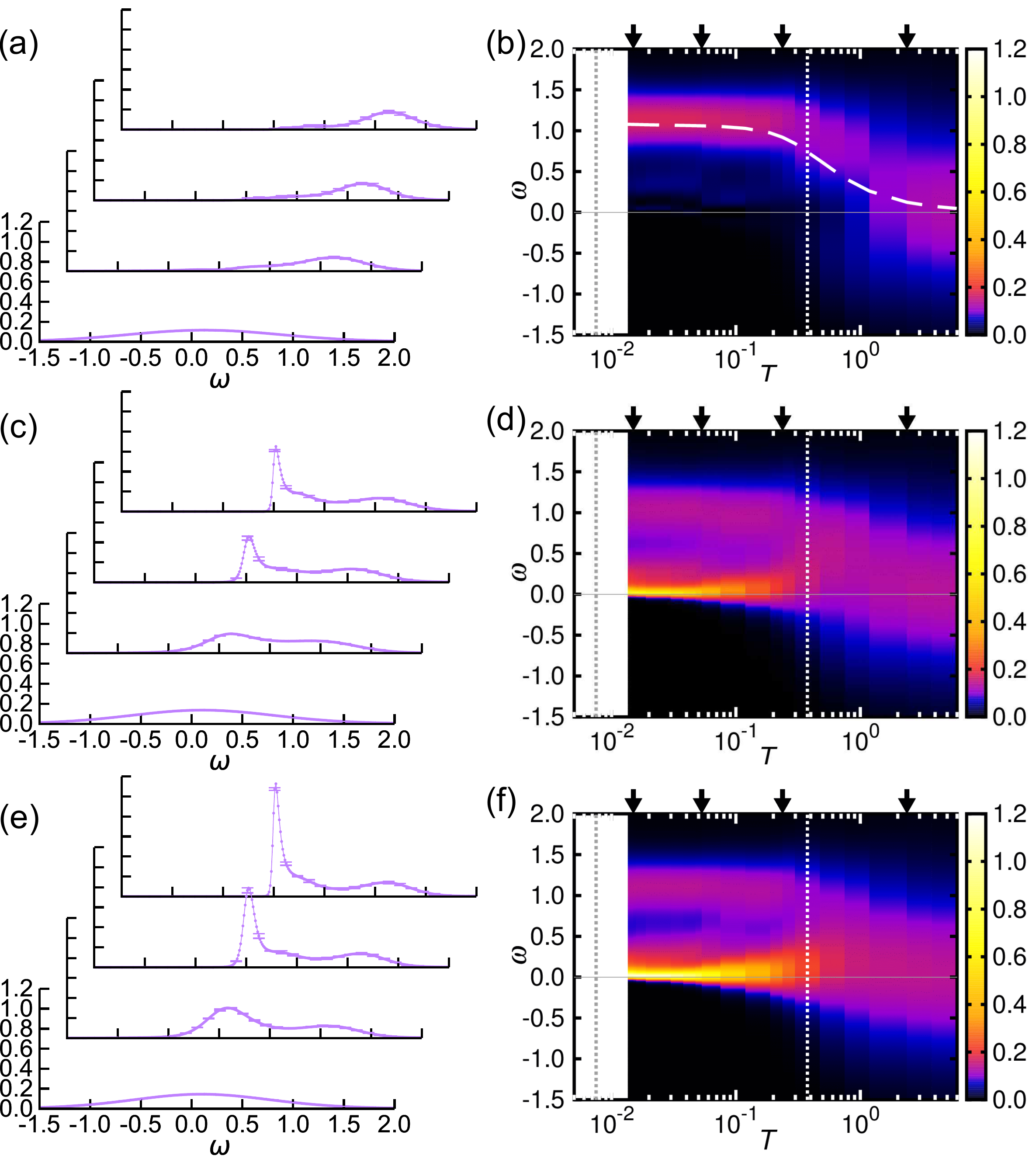}
    \caption{ \label{fig:fig8}
        (a) $S(\Gamma,\omega)$, (c) $S({\rm{K}_1},\omega)$, and (e) $S({\rm{K}_2},\omega)$ for the  AFM case with $\alpha=1.2$ at several $T$.
    The corresponding contour plots in the $T$-$\omega$ plane are shown in (b)(d)(f).
    The notations are common to those in Fig.~\ref{fig:fig5}.
    }
\end{figure}

In order to show the $T$ dependences of $S(\mathbf{q}, \omega)$ more explicitly, we present in Figs.~\ref{fig:fig5}-\ref{fig:fig8} the $T$-$\omega$ plot of $S(\mathbf{q}, \omega)$ at $\mathbf{q} = \Gamma$, ${\rm{K}}_1$, and K$_2$ with the intensity profiles for the same set of $T$ used in Figs.~\ref{fig:fig3} and \ref{fig:fig4}.
Figure~\ref{fig:fig5} shows the result for the FM case at $\alpha = 0.8$.
The overall weight of $S(\mathbf{q}, \omega)$ shifts from $\omega \sim 0$ to a large-$\omega$ region when the system is cooled down below $T \sim T_{\rm{H}}$. 
Below $T_{\rm{H}}$, quasi-elastic response gradually grows and develops to the $\delta$-function like peak.
The peak intensity in $S(\Gamma, \omega)$ and $S({\rm{K}_2}, \omega)$ is larger than that for $S({\rm{K}_1}, \omega)$, reflecting the anisotropy of the interaction.

Figure~\ref{fig:fig6} shows the corresponding plot for the AFM case at $\alpha = 0.8$. 
In contrast to the FM case, the strong quasi-elastic response is seen for $\mathbf{q} = {\rm{K}}_1$, which develops to the $\delta$-function like peak at low $T$.
We note that the dip and shoulder like structures around $\omega=0$ in the intermediate $T$ for the result at $\mathbf{q} = {\rm{K}}_2$ may be an artifact originating from low precision in the MEM for this AFM case because of the following reason. 
As described in Sec.~\ref{subsec:MEM}, we calculate $S^{p}_{j,j'}(\omega)$ for the NN bonds by subtracting the onsite component $S^{p}_{j,j}(\omega)$ from $S^{p}_{j,j}(\omega) + S^{p}_{j,j'}(\omega)$, both of which are obtained by the MEM. 
In the present case, as both of $S^{z}_{j,j}(\omega)$ and $S^{z}_{j,j}(\omega) + S^{z}_{j,j'}(\omega)$ become large around $\omega = 0$ due to the development of the $\delta$-function like peak, the relative error becomes large for $S({\rm{K}_2}, \omega\sim 0)$, which may lead to artificial structures.

Figures~\ref{fig:fig7} and \ref{fig:fig8} show the results at $\alpha = 1.2$. 
As observed in Figs.~\ref{fig:fig3} and \ref{fig:fig4}, $S(\mathbf{q}, \omega)$ for both the FM and AFM cases behave similarly to those at $\alpha = 1.0$~\cite{Yoshitake2016}. 
In the anisotropic cases, however, the difference between $S({\rm{K}}_1, \omega)$ and $S({\rm{K}}_2, \omega)$ is obvious: the quasi-elastic peak for $S({\rm{K}}_1, \omega)$ is larger (smaller) than that for $S({\rm{K}}_2, \omega)$ in the FM (AFM) case.

As discussed in the previous study~\cite{Yoshitake2016}, there is a relation between the static spin correlation and the average frequency of $S(\Gamma, \omega)$, $\bar{\omega} \equiv \int \omega S(\Gamma, \omega) d\omega / \int S(\Gamma, \omega) d\omega$, originating from the sum rule for $S(\mathbf{q}, \omega)$.
$T$ dependences of $\bar\omega$ are shown by white dashed curves in Figs.~\ref{fig:fig5}(b), \ref{fig:fig6}(b), \ref{fig:fig7}(b), and \ref{fig:fig8}(b). 
In all cases, $\bar{\omega}$ is nearly zero for sufficiently high $T$, but it grows at $T \sim T_{\rm{H}}$ and becomes almost independent of $T$ for $T \lesssim T_{\rm{H}}$.
These $T$ dependences are similar to those of the static spin correlation between the NN sites shown in Fig.~\ref{fig:fig2}.

\subsection{NMR relaxation rate\label{subsec:RelaxationRate}}

\begin{figure}[t]
    \includegraphics[width=0.8\columnwidth,clip]{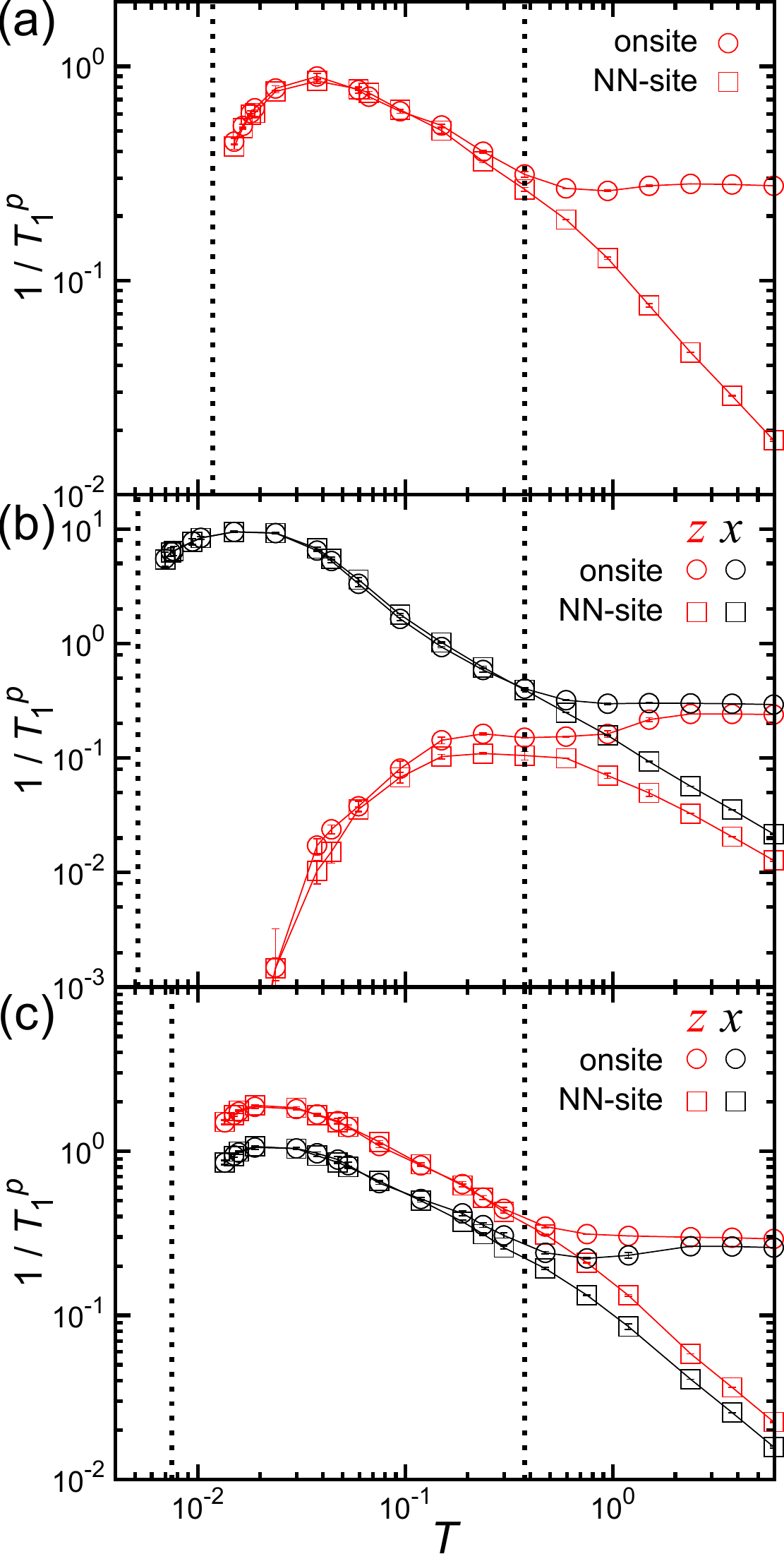}
    \caption{ \label{fig:fig9}
    $T$ dependences of the NMR relaxation rate $1/T^p_1$ ($p=z, x$) at (a) $\alpha=1.0$, (b) $\alpha=0.8$, and (c) $\alpha=1.2$.
    Note that $1/T^z_1 = 1/T^x_1$ for $\alpha=1.0$ and $1/T^x_1 = 1/T^y_1$ for all the cases from the symmetry.
    The vertical dotted lines indicate $T_{\rm L}$ and $T_{\rm H}$ for each $\alpha$. 
    }
\end{figure}

Figure~\ref{fig:fig9} shows the NMR relaxation rate $1/T_1$ obtained by the CDMFT+CTQMC method.
While the results at $\alpha = 1.0$ were presented in the previous study~\cite{Yoshitake2016}, we present them for comparison in Fig.~\ref{fig:fig9}(a).
$1/T_1$ in the magnetic field applied to the $z$ direction, which is denoted by $1/T^{z}_1$, is given by
\begin{align}
1/T^{z}_1 \propto T\sum_{\textbf{q}} |A_\textbf{q}|^2 \frac{{\rm{Im}}\chi^{\perp}(\textbf{q},\omega_0)}{\omega_0},
\label{eq:1/Tz_def}
\end{align}
where $A_\textbf{q}$ is the hyperfine coupling constant, $\chi^{\perp}(\textbf{q},\omega)$ is the dynamical susceptibility for the spin component perpendicular to the magnetic field direction, and $\omega_0$ is the resonance frequency in the NMR measurement.
The dynamical susceptibility $\chi(\textbf{q},\omega)$ is related with the dynamical spin structure factor through the fluctuation-dissipation theorem, as 
\begin{align}
S(\mathbf{q}, \omega) = \frac{1}{\pi (1-e^{-\beta \omega})}{\rm{Im}}\chi(\textbf{q},\omega).
\label{eq:S-chi}
\end{align}
In the NMR experiments, $\omega_0$ is in general negligibly small compared to the typical energy scale of the system, $J$.
Thus, by taking the limit of $\omega_0 \rightarrow 0$ in Eq.~(\ref{eq:1/Tz_def}) and using Eq.~(\ref{eq:S-chi}), we obtain
\begin{align}
1/T^{z}_1 &= a_0 S^{x}_{j,j}(\omega=0) + a_1 S^{y}_{j,j}(\omega=0) \nonumber \\
&+ a_2 S^{x}_{\rm{NN}}(\omega=0) + a_3 S^{y}_{\rm{NN}}(\omega=0),
\end{align}
where the coefficients $a_0$, $a_1$, $a_2$, and $a_3$ are determined by $A_\textbf{q}$.
The similar equations are obtained for $1/T^{x}_1$ and $1/T^{y}_1$ by the cyclic permutation of $x,y,z$ ($1/T_1^x = 1/T_1^y$ for the present cases from the symmetry).
Because $A_\textbf{q}$ depends on the details of the system, we here compute the onsite and NN-site components of $1/T_1$ separately with omitting the coefficients:
the onsite components are calculated as
\begin{align}
1/T_1^z &= S^x_{j,j}(\omega = 0) + S^y_{j,j}(\omega = 0), \\
1/T_1^x &= S^y_{j,j}(\omega = 0) + S^z_{j,j}(\omega = 0),
\end{align}
while the NN-site ones are 
\begin{align}
1/T_1^z &= \pm (S^x_{\rm{NN}}(\omega = 0) + S^y_{\rm{NN}}(\omega = 0) ), 
\label{eq:1/T_1^z}\\
1/T_1^x &= \pm (S^y_{\rm{NN}}(\omega = 0) + S^z_{\rm{NN}}(\omega = 0) ),
\label{eq:1/T_1^x}
\end{align}
where the sign is $+(-)$ for the FM (AFM) case.
We note that, in the anisotropic cases $\alpha \neq 1.0$, the NN-site $1/T_1^x$ is not simply given by the sum in Eq.~(\ref{eq:1/T_1^x}): 
it will be given by a linear combination of $S^y_{\rm{NN}}(\omega = 0)$ and $S^z_{\rm{NN}}(\omega = 0)$ with appropriate coefficients determined by $A_{\mathbf{q}}$.
Such a linear combination, however, can be constructed from our data for Eqs.~(\ref{eq:1/T_1^z}) and (\ref{eq:1/T_1^x}) by noting that $1/T_1^z = 2 S^y_{\rm{NN}}(\omega = 0)$ for $J_x=J_y$. 
Hence, we present the results by Eqs.~(\ref{eq:1/T_1^z}) and (\ref{eq:1/T_1^x}) in Fig.~\ref{fig:fig9} for simplicity.

As shown in Fig.~\ref{fig:fig9}, for all cases, the onsite component of $1/T_1^p$ is nonzero and almost $T$ independent above $T\sim T_{\rm H}$, as expected for the conventional paramagnets~\cite{Moriya1956}. 
On the other hand, the NN-site component is zero in the high-$T$ limit and increases as decreasing $T$.
This behavior corresponds to the development of NN-site static spin correlations shown in Sec.~\ref{subsec:StaticQuantity}, as they have a relation through the sum rule, $\int S_{j,j'}^{p}(\omega) d\omega = \langle S_j^{p}S_{j'}^{p} \rangle$. 

When lowering $T$ below $T_{\rm H}$, $1/T_1^x$ for $\alpha=0.8$ substantially increases, as shown in Fig.~\ref{fig:fig9}(b).
The enhancement is much larger than the case of $\alpha=1.0$ in Fig.~\ref{fig:fig9}(a).
This is due to the evolution of the $\delta$-function like peak in $S^z(\mathbf{q}, \omega)$ discussed in Sec.~\ref{subsec:DSF}.
In contrast, $S^x(\mathbf{q}, \omega)$ and $S^y(\mathbf{q}, \omega)$ do not develop such $\delta$-function like peaks, and hence, $1/T_1^z$ does not show enhancement unlike $1/T_1^x$.
While further decreasing $T$, $1/T_1^x$ shows a peak slightly above $T_{\rm{L}}$. 
The decrease at low $T$ reflects a spin gap originating from the nonzero flux gap in the ground state~\cite{Kitaev2006}.
On the other hand, the onsite and NN-site components of $1/T_1^z$ are both suppressed below $T \sim T_{\rm H}$, after showing a plateau and broad peak, respectively.
The suppression of $1/T_1^z$ is due to an increase of energy cost for a spin flip on the strong $z$ bond under the well-developed static spin correlations between NN sites in this $T$ range.
Actually, the energy cost is represented by the average frequency of $S_{j,j}^{x}(\omega)$ as there is a relation 
\begin{align}
\bar{\omega}^{x}_{\rm{onsite}}
&=\frac{\int \omega S_{j,j}^{x}(\omega) d\omega}{ \int S_{j,j}^{x}(\omega) d\omega } \nonumber \\
&= \frac{\sum_{m,n} e^{-\beta E_n} (E_m - E_n) |\langle m| S^x_j |n\rangle|^2}{\frac{1}{4}\sum_{n} e^{-\beta E_n} }.
\end{align}
On the other hand, $\bar{\omega}^{x}_{\rm{onsite}}$ is also written as $\bar{\omega}^{x}_{\rm{onsite}}= (J_y\langle S_j^y S_{j'}^y \rangle_{\rm{NN}} + J_z\langle S_j^z S_{j'}^z \rangle_{\rm{NN}})/2$ by the sum rule~\cite{Yoshitake2016}.
Thus, the energy cost becomes large below $T_{\rm{H}}$ according to the growth of $\langle S_j^z S_{j'}^z \rangle_{\rm{NN}}$.

In contrast, as shown in Fig.~\ref{fig:fig9}(c), $T$ dependence of $1/T_1^p$ at $\alpha=1.2$ is similar to that at $\alpha=1.0$ in Fig.~\ref{fig:fig9}(a).
Both $1/T_1^x$ and $1/T_1^z$ increase below $T_{\rm{H}}$ while decreasing $T$, in contrast to the case with $\alpha=0.8$.
For $\alpha=1.2$, however, $1/T_1^z$ is larger than $1/T_1^x$, reflecting the stronger interactions on the $x$ and $y$ bonds than the $z$ bond.
On further decreasing $T$, $1/T_1^p$ at $\alpha=1.2$ also shows the peak structure slightly above $T_{\rm{L}}$ and then decreases, as expected from the finite flux gap in the ground state.

Although the system is described by free Majorana fermions coupled to localized gauge fluxes, the NMR relaxation rate does not obey the Korringa law, $1/(T_1T) \sim {\rm constant}$, which is expected for free fermion systems.
This is natural because the spin-flip excitation in the NMR process is a composite of both itinerant matter fermions and localized gauge fluxes. 
Nonetheless, for comparison to forth-coming experiments, we plot the Korringa ratio as a function of $T$ in Appendix~\ref{app:Korringa}.

\subsection{Magnetic susceptibility\label{subsec:Susceptibility}}

\begin{figure}[t]
    \includegraphics[width=0.84\columnwidth,clip]{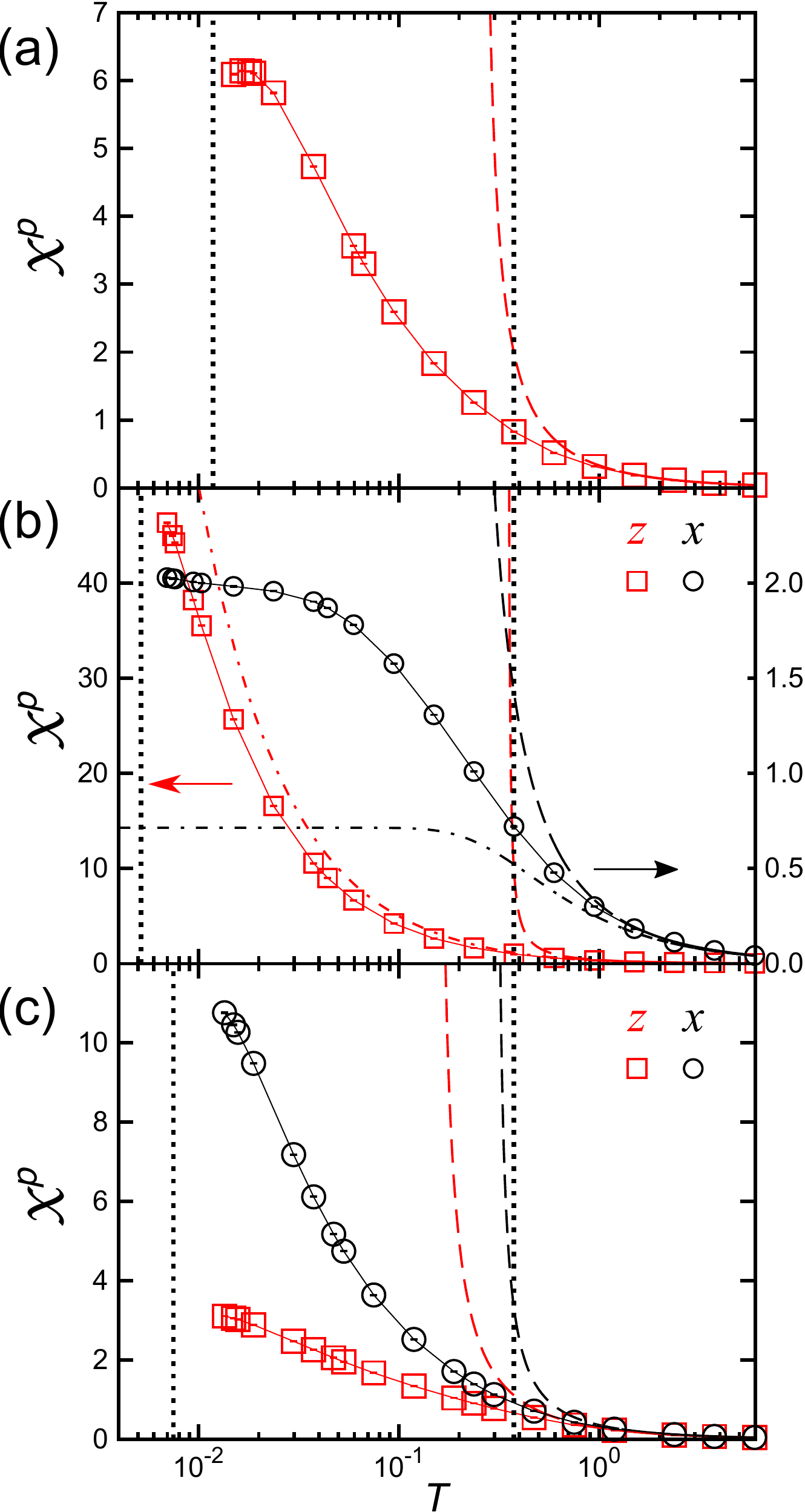}
    \caption{ \label{fig:fig10}
    $T$ dependences of the magnetic susceptibility $\chi^p$ ($p=z, x$) at (a) $\alpha = 1.0$, (b) $\alpha = 0.8$, and (c) $\alpha = 1.2$ for the FM case.
    The dashed curves represent $\chi^p_{\rm{CW}}$ in Eq.~(\ref{eq:chi_CW}).
    The red and black dashed-dotted curves in (b) represent $\chi^p_{\rm dimer}$ for $p=z$ [Eq.~(\ref{eq:chi_dimer^z})] and $p=x$ [Eq.~(\ref{eq:chi_dimer^x})], respectively.
    Note that $\chi^z = \chi^x$ for $\alpha=1.0$ and $\chi^x=\chi^y$ for all the cases from the symmetry.
    The vertical dotted lines indicate $T_{\rm L}$ and $T_{\rm H}$ for each $\alpha$. 
    }
\end{figure}

\begin{figure}[t]
    \includegraphics[width=0.84\columnwidth,clip]{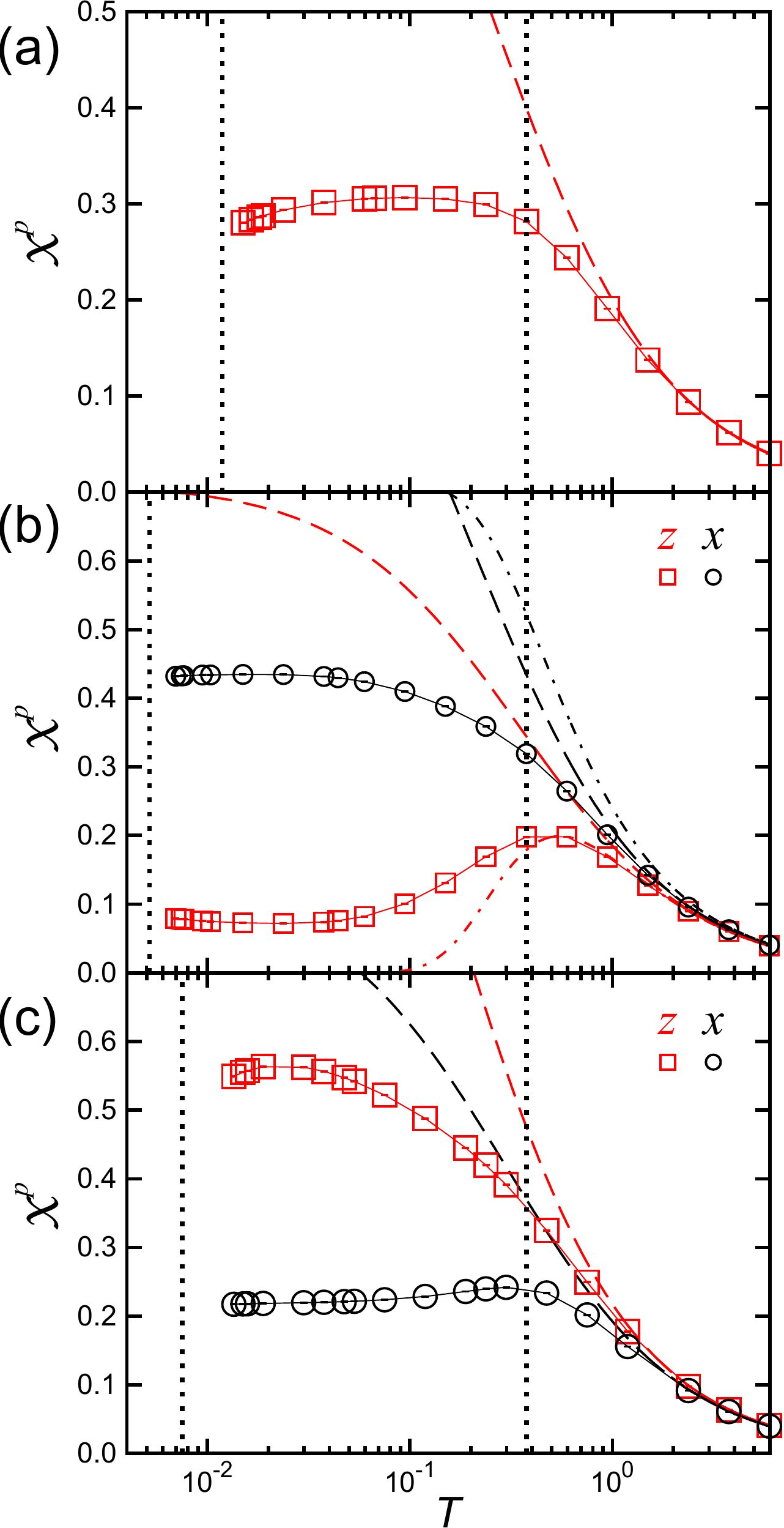}
    \caption{ \label{fig:fig11}
    $T$ dependences of the magnetic susceptibility $\chi^p$ ($p=z, x$) at (a) $\alpha = 1.0$, (b) $\alpha = 0.8$, and (c) $\alpha = 1.2$ for the AFM case.
    The notations are common to those in Fig.~\ref{fig:fig10}.
    }
\end{figure}

Figures~\ref{fig:fig10} and \ref{fig:fig11} show the $T$ dependences of the magnetic susceptibility $\chi^p$ for the FM and AFM cases, respectively.
$\chi^p$ at $\alpha = 1.0$ presented in the previous study~\cite{Yoshitake2016}, are also presented in Fig.~\ref{fig:fig10}(a) and Fig.~\ref{fig:fig11}(a) for comparison.
$\chi^p$ is calculated from the imaginary-time spin correlations as
\begin{align}
\chi^p = \frac{1}{N}
\sum_{j,j'} \int_0^\beta d\tau\langle S^p_j (\tau) S^p_{j'}\rangle.
\end{align}
Note that this is obtained without the MEM.
In all the cases, at sufficiently high $T$ compared to the dominant $J_p$, $J_p^{\rm max} = {\rm max}\{J_p\}$, $\chi^p$ obeys the Curie-Weiss law,
\begin{align}
\chi^p_{\rm{CW}} = \frac{1}{4T-J_p^{\rm max}}, 
\label{eq:chi_CW}
\end{align}
which is obtained by the standard mean-field approximation in the original spin representation.
While decreasing $T$, $\chi^p$ shows a deviation from $\chi^p_{\rm{CW}}$ below $T\sim  J_p^{\rm max}$.

Among the results, $\chi^x$ for the FM case and $\chi^z$ for the AFM case at $\alpha=0.8$ show
peculiar $T$ dependences at low $T$.
The former largely deviates from the Curie-Weiss behavior and saturates to a small nonzero value, as shown in Fig.~\ref{fig:fig10}(b)~\cite{note1}. 
Meanwhile, the latter shows a broad hump at $T\sim T_{\rm H}$ and decreases as lowering $T$, as shown in Fig.~\ref{fig:fig11}(b).
These $T$ dependences are qualitatively understood by considering a two-site dimer model on the $z$ bond obtained by setting $J_x = J_y = 0$. 
The dimer model gives the analytical forms for the magnetic susceptibility as
\begin{align} 
\chi_{\rm dimer}^z &= \frac{\beta}{2} \frac{ \exp(\beta J_z/4)}{\exp(\beta J_z/4)+\exp(-\beta J_z/4)},
\label{eq:chi_dimer^z}\\
\chi_{\rm dimer}^x &= \frac{1}{J_z}\tanh\left(\beta \frac{J_z}{4}\right).
\label{eq:chi_dimer^x}
\end{align}
The results are plotted by the dashed-dotted curves in Figs.~\ref{fig:fig10}(b) and \ref{fig:fig11}(b).
$\chi_{\rm dimer}^x$ for the FM case almost saturates around $T\sim J_z/4$, as the dominant $J_z$ interaction suppresses the magnetization in the $x$ direction.
This accounts for the behavior of $\chi^x$ in Fig.~\ref{fig:fig10}(b) qualitatively. 
Meanwhile, $\chi_{\rm dimer}^z$ also well reproduces a hump at $T\sim 0.5$ in $\chi^z$ for the AFM case in Fig.~\ref{fig:fig11}(b); 
$\chi^z$ remains nonzero down to low $T$ as nonzero $J_x$ and $J_y$ smear out the dimer gap.

In the case of $\alpha=1.2$, $T$ dependences of $\chi^p$ shown in Figs.~\ref{fig:fig10}(c) and \ref{fig:fig11}(c) are similar to those for $\alpha=1.0$ in the previous study~\cite{Yoshitake2016} replotted in Figs.~\ref{fig:fig10}(a) and \ref{fig:fig11}(a), respectively; 
on decreasing $T$, $\chi^p$ continues to increase down to $T\sim T_{\rm{L}}$ in the FM case, whereas $\chi^p$ shows broad peak at a higher $T$ in the AFM case.
The effect of anisotropic $J_p$, however, is clearly observed: 
the stronger interactions on the $x,y$ bonds than the $z$ bond result in larger (smaller) $\chi^x$ than $\chi^z$ in the FM (AFM) case.
In addition, the temperature of the broad peak of $\chi^z$ ($\chi^x$) in the AFM case shifts to a lower (higher) $T$ than that for $\alpha = 1.0$.

\section{Discussion\label{sec:Discussion}}

As pointed out in the previous study for the isotropic case by the authors~\cite{Yoshitake2016} and confirmed also for the anisotropic cases in the present study, a remarkable feature in the Kitaev model is the dichotomy between the dynamical and static spin correlations; 
namely, the NMR relaxation rate $1/T_1^p$ and the magnetic susceptibility, both of which reflect the dynamical spin correlations, show substantial $T$ dependences below $T_{\rm H}$ (Figs.~\ref{fig:fig9}-\ref{fig:fig11}), even though the static spin correlations $\langle S_j^p S_{j'}^p \rangle_{\rm{NN}}$ almost saturate to the $T=0$ values (Fig.~\ref{fig:fig2}).
The dichotomy is unconventional behavior hardly seen in conventional insulating magnets.
This might be a signature of the fractionalization of quantum spins, as $T_{\rm H}$ is the temperature where the fractionalization sets in as indicated in the specific heat and entropy~\cite{Nasu2015}.

To examine the dichotomy in more detail, we calculate the $T$ dependences of $1/T_1^p$ and $\langle S_j^p S_{j'}^p \rangle_{\rm{NN}}$ for two extreme cases by assuming the configuration of $\{\eta\}$ by hand.
One is the flux-free state with all $\eta_r=+1$, which is realized in the ground state. 
The other is the state with completely random $\{\eta \}$, corresponding to the high-$T$ limit.
For this purpose, we regard a single $z$ bond $r_0$ as the cluster in CDMFT,
and take $P(\eta_{r_0}=1) = 1$ and $P(\eta_{r_0}=-1) = 0$ for the former uniform state, while $P(\eta_{r_0}=1) = P(\eta_{r_0}=-1) = 1/2$ for the latter random state, in Eq.~(\ref{eq:localG}) of the self-consistent equation of CDMFT~\cite{note2}. 

\begin{figure*}[t]
    \includegraphics[width=2\columnwidth,clip]{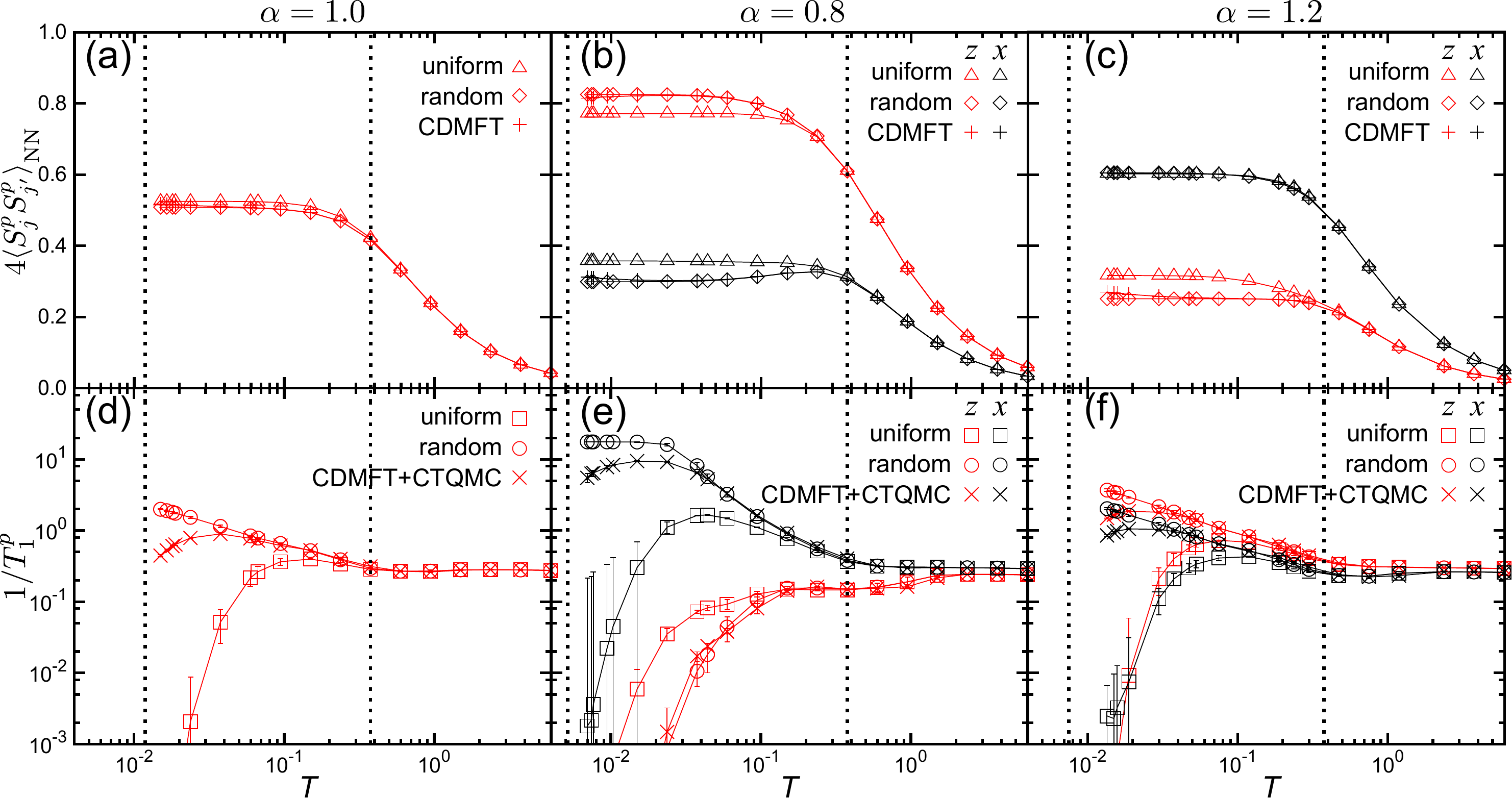}
    \caption{ \label{fig:fig12}
    (a)(b)(c) 4$\langle S_j^p S_{j'}^p \rangle_{\rm{NN}}$ for the FM case and (d)(e)(f) the onsite components of $1/T^p_1$ ($p=z, x$) calculated by setting all $\eta = 1$ (uniform) and all $\eta$ being random (random) in the CDMFT calculations: (a)(d) $\alpha=1.0$, (b)(e) $\alpha=0.8$, and (c)(f) $\alpha=1.2$.
    In (a) and (d), $\langle S_j^p S_{j'}^p \rangle_{\rm{NN}}$ and $1/T^p_1$ are equivalent for $p=x,z$.
    For comparison, we plot the data in Figs.~\ref{fig:fig2} (CDMFT) and \ref{fig:fig9} (CDMFT+CTQMC). 
    The vertical dotted lines represent $T_{\rm L}$ and $T_{\rm H}$ for each $\alpha$. 
    }
\end{figure*}

Figure~\ref{fig:fig12} shows the results. 
In all cases, $\langle S_j^p S_{j'}^p \rangle_{\rm{NN}}$ for both uniform and random $\{\eta\}$ shows almost similar $T$ dependence to the CDMFT results, as shown in Figs.~\ref{fig:fig12}(a)-\ref{fig:fig12}(c).
However, $1/T_1^p$ exhibits considerably different $T$ dependence.
For instance, in the isotropic case with $\alpha = 1.0$, although $1/T_1$ is almost $T$ independent for $T>T_{\rm H}$ for both uniform and random $\{\eta\}$ similar to the result by the CDMFT+CTQMC method in Ref.~\cite{Yoshitake2016},
it shows different behavior below $T_{\rm H}$ between the two cases, as shown in Fig.~\ref{fig:fig12}(d). 
For the case with uniform $\{\eta\}$, $1/T_1$ decreases to zero after showing a small hump. 
The suppression at low $T$ reflects the flux gap $\Delta\simeq 0.065J$ in the flux-free state~\cite{Kitaev2006, Knolle2014}.
On the other hand, for the case with random $\{\eta\}$, $1/T_1$ monotonically increases while decreasing $T$ in the calculated $T$ range.
Similar $T$ dependences of $1/T_1^p$ are obtained for $1/T_1^x$ at $\alpha=0.8$ and $1/T_1^{x,z}$ at $\alpha=1.2$, as shown in Figs.~\ref{fig:fig12}(e) and \ref{fig:fig12}(f), respectively.
We note that $1/T^z$ for $\alpha=0.8$ behaves differently; we will comment on this point in the end of this section.

The results clearly indicate that the peculiar $T$ dependences of $1/T_1$ found in the CDMFT+CTQMC results are closely related with fluctuations of the gauge fluxes $\{\eta \}$ composed of localized Majorana fermions $\{\bar{c}\}$ emergent from the spin fractionalization.
As seen in the equal-time spin correlations shown in Figs.~\ref{fig:fig12}(a)-\ref{fig:fig12}(c), itinerant matter fermions develop their kinetic energy to the saturation at $T \sim T_{\rm H}$ (the equal-time spin correlations correspond to the kinetic energy of matter fermions).
Due to the fractionalization, however, the localized gauge fluxes are still disordered even below $T_{\rm H}$~\cite{Nasu2015}, which results in the enhancement of $1/T_1$, as indicated in Figs.~\ref{fig:fig12}(d)-\ref{fig:fig12}(f).
When approaching $T_{\rm L}$, $\{\eta \}$ are aligned in a coherent manner~\cite{Nasu2015}, and hence, $1/T_1$ is rapidly suppressed at $T\sim T_{\rm L}$. 
Thus, the $T$ dependence of $1/T_1$ is qualitatively explained by the crossover from that for the random $\{\eta \}$ to the fully-aligned $\{\eta \}$ while decreasing $T$. 
The crossover occurs well below $T_{\rm H}$ and close to $T_{\rm L}$.
Of course, as the original quantum spin is a composite of itinerant matter fermions and localized gauge fluxes, the spin-flip dynamics is a composite excitation.
Nevertheless, our results indicate that the peculiar $T$ dependence of the NMR relaxation rate as well as the magnetic susceptibility is dominated by the emergent gauge fluxes from the fractionalization.

As noted above, $1/T_1^z$ for $\alpha=0.8$ behaves differently from others: $1/T_1^z$ for the random $\{\eta\}$ is smaller than that for the uniform $\{\eta\}$ at low $T$, as shown in Fig.~\ref{fig:fig12}(e). 
This is presumably because of the peculiar $T$ dependence of the density of states (DOS) for the itinerant matter fermions at $\alpha=0.8$.
In the gapless QSL region for $0.75 \leq \alpha \leq 1.5$ but close to the gapless-gapful boundary at $\alpha=0.75$, the DOS opens a gap as $\{\eta\}$ are thermally disordered by raising $T$~\cite{Nasu2015}.
Thus, the DOS for matter fermions is gapless for the uniform $\{\eta\}$, while gapped for the random $\{\eta\}$. 
As spin excitations by $S_j^x$ and $S_j^y$ are composite excitations of both itinerant matter fermions and localized gauge fluxes, the gap in the DOS for matter fermions suppresses $1/T_1^z$ for the random case compared to the uniform one. 
Since $\{\eta\}$ are aligned uniformly below $T_{\rm L}$, we expect that $1/T_1^z$ shows an abrupt increase while decreasing $T$ through $T_{\rm L}$. 
This indicates that while a rapid change of $1/T_1$ when approaching $T_{\rm L}$ is yielded by the coherent alignment of $\{\eta\}$, either increase or decrease of $1/T_1$ at $T_{\rm L}$ may be affected by the itinerant matter fermions.

\section{Summary}

To summarize, we have presented numerical results for spin dynamics of the Kitaev model with the anisotropy in the bond-dependent coupling constants.
We calculated the experimentally-measurable quantities, the dynamical spin structure factor $S(\bf{q},\omega)$, the NMR relaxation rate $1/T_1$, and the magnetic susceptibility $\chi$, in the wide $T$ range including the peculiar paramagnetic region where quantum spins are fractionalized.
The results have been obtained by the Majorana CDMFT+CTQMC method, which were developed by the authors previously~\cite{Yoshitake2016}; 
we gave detailed descriptions of the method, including the MEM for analytical continuation.
We also confirmed the Majorana CDMFT is precise enough in the range of $T$ and anisotropy that we investigated in the present study.

We found that the Kitaev model exhibits unconventional behaviors in spin dynamics in the finite-$T$ paramagnetic state in proximity to the QSL ground state. 
The prominent feature is the dichotomy between static and dynamical spin correlations as a consequence  of the spin fractionalization.
The dichotomy appears clearly in the increase of $1/T_1$ below $T_{\rm H}$ where the fractionalization sets in, despite the saturation of static correlations.
Similar behavior was also seen in the isotropic case in the previous study~\cite{Yoshitake2016}. 
Our results suggest that the dichotomy is found universally in the fractionalized paramagnetic region irrespective of the anisotropy in the system.

On the other hand, we also clarified interesting behaviors that depend on the anisotropy at low $T$. 
When one of the three bond-dependent interactions is stronger than the other two, the spin dynamics shows peculiar $T$ and energy dependences distinct from those in the isotropic coupling case as follows.
As lowering $T$, $S(\bf{q},\omega)$ develops a $\delta$-function like peak, which is well separated from the incoherent continuum. 
$1/T_1$ monotonically decreases in the spin component for the stronger bond. 
$\chi$ increases and saturates to a nonzero value for the spin component for the weaker bonds, while it shows hump and then decreases for the stronger-bond component in the antiferromagnetic case. 
We also showed that the peculiar $T$ dependences of $\chi$ are qualitatively explained by the two-site dimer model.
In contrast, when the anisotropy is opposite, i.e., when the two types of bonds become stronger, the results are qualitatively unchanged from those for the isotropic case, while the effect of anisotropy is obvious in the $\bf{q}$ dependence in $S(\bf{q},\omega)$ and the different components in $1/T_1$ and $\chi$.

Our results will stimulate further experimental and theoretical analyses of candidate materials for the Kitaev QSLs. 
As most of the materials are assumed to be anisotropic in the exchange constants~\cite{Choi2012, Johnson2015, Yamaji2014, Winter2016}, our results will be helpful for understanding of unusual behaviors in the real compounds. 
We emphasize that our numerical data obtained by the Majorana CDMFT+CTQMC method are quantitatively reliable in the calculated paramagnetic regime, as the cluster approximation and the analytic continuation are both well controlled. 
Although there are residual interactions in addition to the Kitaev-type ones in real materials, our results provide good references in the limit of the pure Kitaev model for interpreting the role of the additional interactions.

While we have calculated dynamical quantities of the Kitaev model in the wide $T$ range, the calculations were limited above $T_{\rm{L}}$ due to the phase transition which is artifact of the mean-field nature of CDMFT.
It is necessary to develop more sophisticated method to study the dynamical properties below $T_{\rm L}$. 
The low-$T$ spin dynamics will be interesting, in particular, for extensions of the Kitaev model to three-dimensional lattices, such as hyperhoneycomb and hyperoctagon lattices~\cite{Hermanns2014}.
In the three-dimensional cases, in general, the Kitaev models may cause a finite-$T$ phase transition between the paramagnetic and QSL phases.
Indeed, such an exotic transition was found for the hyperhoneycomb Kitaev model~\cite{Nasu2014}.
The phase transition is triggered by the confinement and deconfinement of emergent loops composed of excited fluxes~\cite{Nasu2014}.
This is a topological phase transition that cannot be described by a local order parameter.
Although it is expected that dynamical quantities exhibit peculiar behavior associated with the topological phase transition, the CDMFT is not able to describe such a transition.
Thus, with bearing the fact in mind that there are some candidates for the three-dimensional Kitaev QSLs~\cite{Okamoto2007,Kuriyama2010,Modic2014, Takayama2015,Kimchi2014} the calculation of dynamical quantities in all $T$ range beyond the CDMFT is an interesting challenge left for future works.

\appendix \section{Cluster size dependence\label{app:SizeDep}}

In the CDMFT, we replace the lattice model to the impurity model with a finite-size cluster. The CDMFT becomes exact in the limit of infinite size cluster.
Although the cluster size dependence was examined for the isotropic case with $\alpha=1.0$ in Supplemental Material for the previous study~\cite{Yoshitake2016}, here we present the cluster size dependences of $\chi^p$ and $1/T_1^p$ for $\alpha=0.8$ and $1.2$ in comparison with the $\alpha=1.0$ case.
As the onsite and NN-site components of $1/T_1^p$ shows almost the same $T$ dependences below $T_{\rm{H}}$ (see Fig.~\ref{fig:fig9}), we present only the onsite one.

\begin{figure}[t]
    \includegraphics[width=\columnwidth,clip]{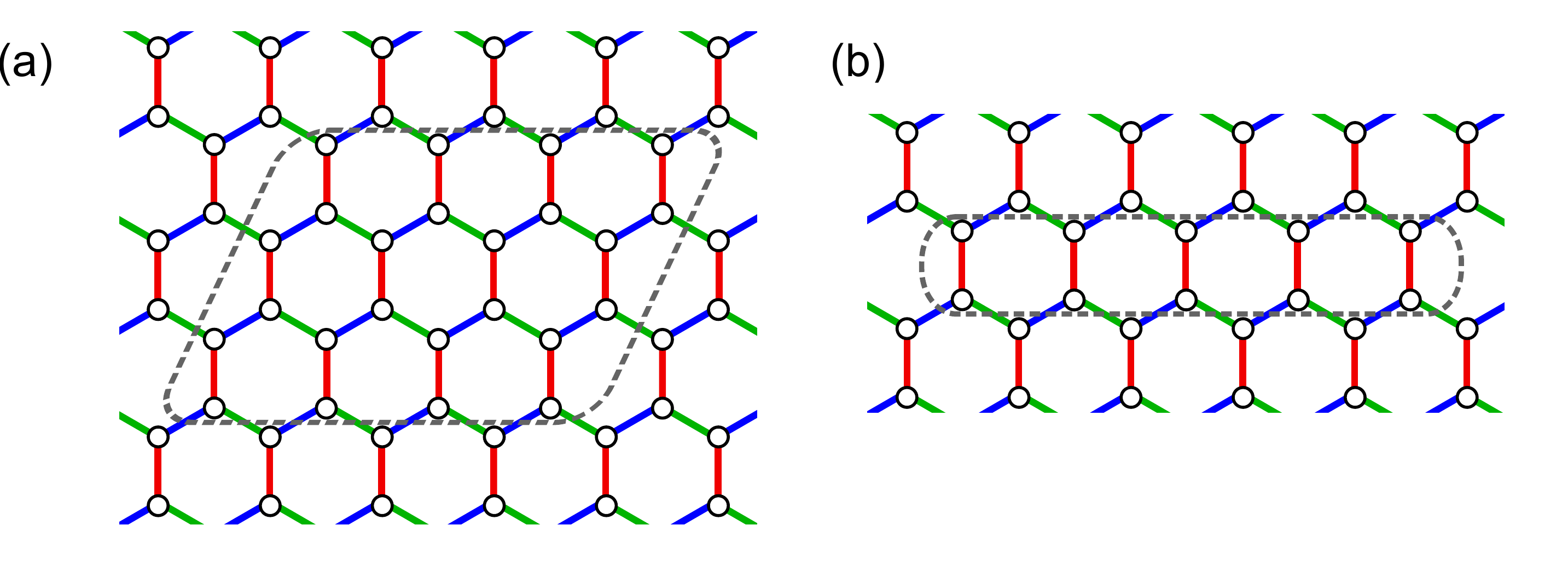}
    \caption{ \label{fig:fig13}
    Schematic pictures of the different types of clusters used in the benchmark of CDMFT.
    The color of the bonds are common to Fig.~\ref{fig:fig1}(a).
    }
\end{figure}

\begin{figure*}[t]
   \includegraphics[width=2.05\columnwidth,clip]{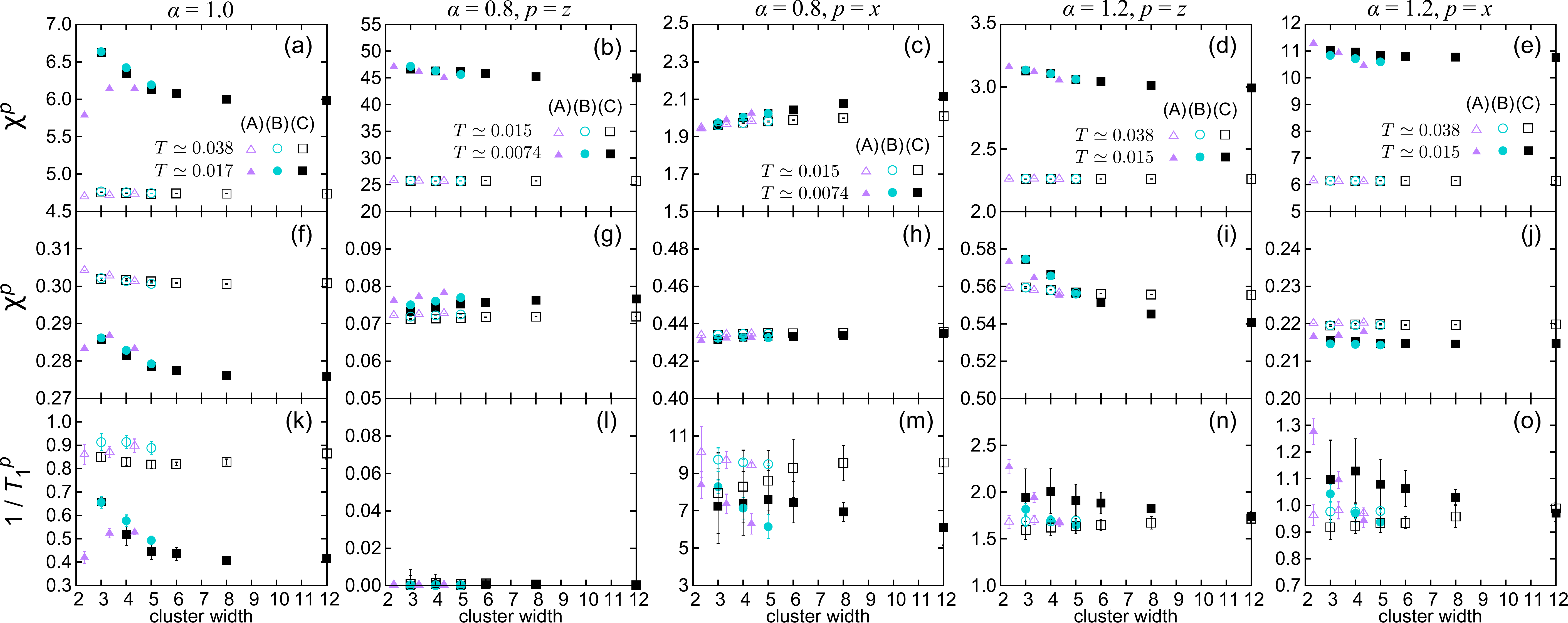}
    \caption{ \label{fig:fig14}
    Cluster-size dependences of the magnetic susceptibility $\chi^p$ for (a)-(e) the FM case and (f)-(j) the AFM case, and (k)-(o) the onsite component of the NMR relaxation rate $1/T^p_1$: (a)(f)(k) $\alpha=1.0$, (b)(c)(g)(h)(l)(m) $\alpha=0.8$, and (d)(e)(i)(j)(n)(o) $\alpha=1.2$. The data for two different $T$ are plotted in each case.
    In (a)(f)(k), the data are common to $p=z$ and $x$.
    Calculations are performed for the cluster series denoted in (A) Fig.~\ref{fig:fig1}(a), (B) Fig.~\ref{fig:fig13}(a), and (C) Fig.~\ref{fig:fig13}(b).
    Symbols in (a)-(e) are common for the same parameters in (f)-(o).
    }
\end{figure*}

Figure~\ref{fig:fig14} shows the cluster size dependence of $\chi^p$ and $1/T_1^p$ obtained by the CDMFT+CTQMC calculations for three different types of clusters shown in Figs.~\ref{fig:fig1}(a), \ref{fig:fig13}(a), and \ref{fig:fig13}(b).
In each type, we change the cluster sizes in the width in the $xy$-chain direction while keeping that in the $z$-bond direction.
This is because the width in the $xy$-chain direction is rather relevant compared to that in the $z$-bond direction in the present CDMFT, presumably due to the Majorana representation based on the Jordan-Wigner transformation along the $xy$ chains.
Hereafter, we define the size of the cluster by the average width in the $xy$-chain direction: for instance, $4.3$ for the cluster in Fig.~\ref{fig:fig1}(a), while $4$ and $5$ for Figs.~\ref{fig:fig13}(a) and \ref{fig:fig13}(b), respectively.

As shown in Figs.~\ref{fig:fig14}(a)-\ref{fig:fig14}(j), the CDMFT+CTQMC results for $\chi^p$ show quick convergence with respect to the cluster width for all the cluster types.
Even close to the artificial critical temperature $\tilde{T}_{\rm c}$, the results for the width larger than $4$ are almost convergent to the large width limit for all types of the clusters: the remnant relative errors are $\lesssim 5$\%.
Note that 
$\tilde{T}_{\rm c}\sim 0.014$ for $\alpha = 1.0$,
$\tilde{T}_{\rm c}\sim 0.0063$ for $\alpha = 0.8$, and
$\tilde{T}_{\rm c}\sim 0.013$ for $\alpha=1.2$
(for the rotated lattice coordinate used to calculate $\langle S^{p}_{j}(\tau)S^{p}_{j'} \rangle$ for $p=x, y$,
$\tilde{T}_{\rm c}$ becomes slightly lower: $\tilde{T}_{\rm c}\sim 0.0052$ for $\alpha = 0.8$ and $\tilde{T}_{\rm c} \sim 0.0094$ for $\alpha=1.2$).

On the other hand, as shown in Figs.~\ref{fig:fig14}(k)-\ref{fig:fig14}(o), the cluster-size dependences of $1/T_1$ remains up to relatively higher $T$ than $\chi^p$.
But the remnant relative errors are $\lesssim 10$\% for the cluster width larger than $4$, which are sufficiently small to observe the characteristic $T$ dependences of $1/T_1$ as shown in Figs.~\ref{fig:fig9}.

\section{Accuracy of the maximum entropy method\label{app:MEM}}

In the CDMFT+CTQMC calculations, we calculate $S^p_{j,j'}(\omega)$ from $\langle S^{p}_{j}(\tau)S^{p}_{j'} \rangle$ by the MEM as described in Sec.~\ref{subsec:MEM}.
In this Appendix, we examine the accuracy of the MEM in the limit of decoupled one-dimensional chains, i.e., $\alpha=1.5$ ($J_z=0$), where $S^p_{j,j'}(\omega)$ can be obtained directly without the MEM.
We also examine the accuracy by comparing $S^p_{j,j'}(\omega)$ at sufficiently low-$T$ with the analytical solution in the ground state.

First, we show the comparison in the limit of decoupled one-dimensional chains, i.e., $\alpha=1.5$ ($J_z=0$). 
In this limit, the Kitaev Hamiltonian in Eq.~(\ref{eq:Hamiltonian0}) is written only by itinerant matter fermions $\{c\}$, in the form of Eq.~(\ref{eq:Hamiltonian1}) with $J_z=0$.
In this noninteracting problem, following Ref.~\cite{Derzhko2000}, we can calculate $S^x_{j,j'}(\omega)$ by considering the real-time evolution (RTE) of $\langle S^{x}_{j}(t)S^{x}_{j'} \rangle$, instead of the imaginary-time correlation $\langle S^{x}_{j}(\tau)S^{x}_{j'} \rangle$, as
 \begin{align}
S^x_{j,j'}(\omega) = \int_{-\infty}^\infty dt e^{i\omega t -\epsilon |t|}\langle S^{x}_{j}(t)S^{x}_{j'} \rangle. 
\label{eq:SOmega}
\end{align}
We call this method as the RTE in the following.
In the RTE calculations, we consider an $xy$ chain with 600 sites under the open boundary condition and take a sufficiently small $\epsilon = 0.04$ in Eq.~(\ref{eq:SOmega}).

On the other hand, $S^z_{j,j'}(\omega)$ has a nonzero value only for the onsite component, which is given by $4\langle S^{z}_{j}(\tau)S^{z}_{j} \rangle = \langle c_{j}(\tau)c_{j} \rangle$.
Hence, $S^z_{j,j}(\omega)$ is obtained as 
\begin{align}
S^z_{j,j}(\omega) = \frac{1}{2(1+e^{-\beta\omega})}D(\omega),
\end{align}
where $D(\omega)$ is the DOS for itinerant matter fermions in the one-dimensional limit:
\begin{align}
D(\omega) = \frac{1}{\pi\sqrt{1.5^2-\omega^2}}. 
\end{align}
We call this method to estimate $S^z_{j,j}(\omega)$ the exact-DOS in the following.

\begin{figure}[t]
    \includegraphics[width=\columnwidth,clip]{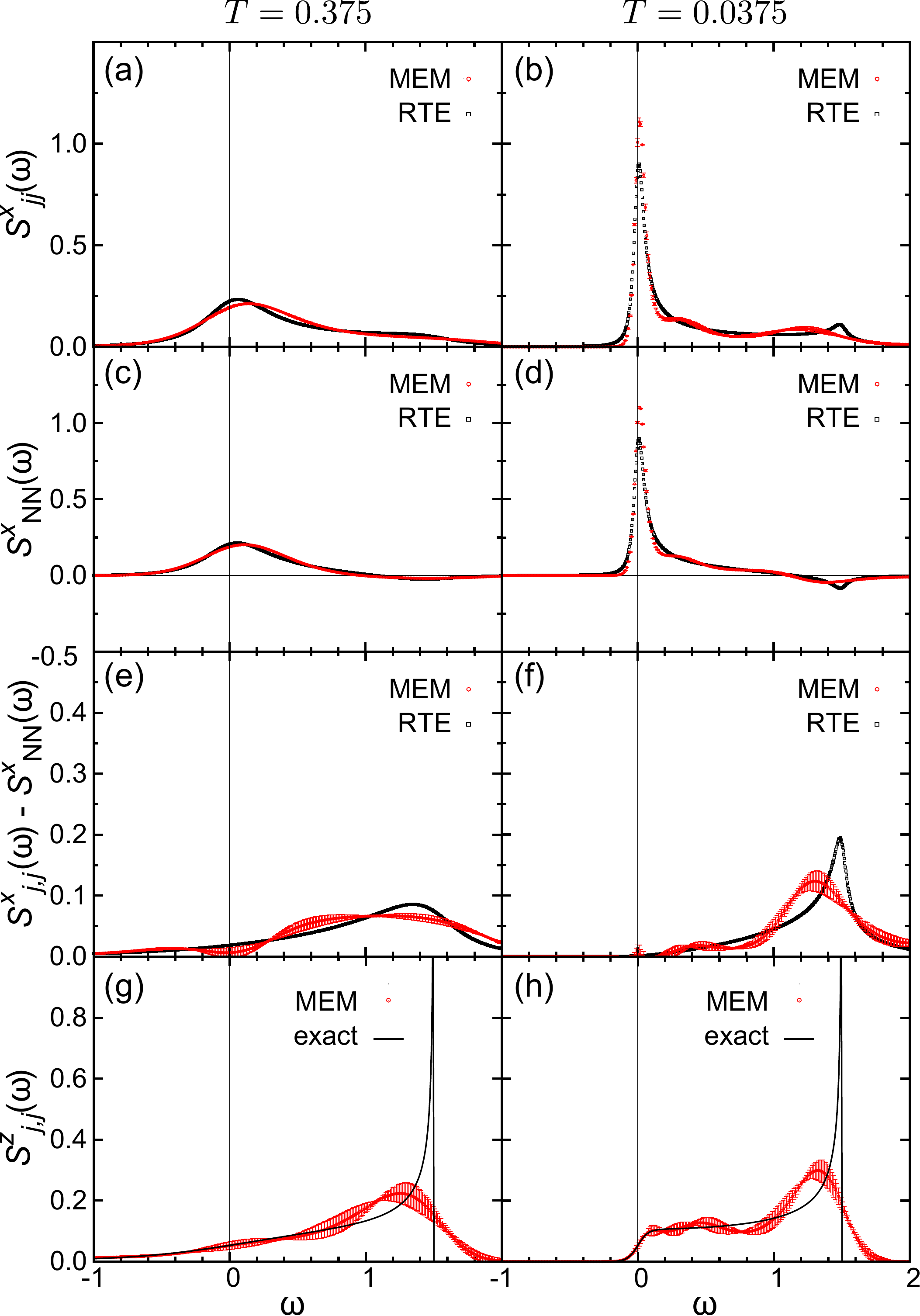}
    \caption{ \label{fig:fig15}
    Comparison between the MEM, RTE, and exact-DOS results for (a)(b) $S^x_{j,j}(\omega)$, (c)(d) $S^x_{\rm{NN}}(\omega)$, (e)(f) $S^x_{j,j}(\omega) - S^x_{\rm{NN}}(\omega)$, and (g)(h) $S^z_{j,j}(\omega)$ at (a)(c)(e)(g) $T=0.375$ and (b)(d)(f)(h) $T=0.0375$.
    }
\end{figure}

Figure~\ref{fig:fig15} shows the results of $S^p_{j,j'}(\omega)$ obtained by the MEM, RTE, and exact-DOS methods for the FM case with $\alpha=1.5$ ($J_x=J_y=1.5$ and $J_z=0$).
We present both onsite and NN-site components for $S^x_{j,j'}(\omega)$, while only the onsite one for $S^z_{j,j'}(\omega)$.
We find that overall $\omega$ dependence of $S^p_{j,j'}(\omega)$ is well reproduced by the MEM. 
In particular, the agreement is excellent in the low $\omega$ region; 
the growth of $S^x_{j,j'}(\omega = 0)$ on decreasing $T$, which contributes to $1/T_1$, is well reproduced by the MEM.
On the other hand, the relatively sharp structures at $\omega\sim 1.5$ are blurred in the MEM results for both $p=x$ and $z$, presumably because $\langle S^{p}_{j}(\tau)S^{p}_{j'} \rangle$ is more insensitive to $S^p_{j,j'}(\omega)$ in the larger $\omega$ region.
Nevertheless, as shown in Figs.~\ref{fig:fig15}(e) and \ref{fig:fig15}(f), the MEM results reproduce the broad incoherent peak of $S^x_{j,j}(\omega) - S^x_{\rm{NN}}(\omega)$.

\begin{figure}[t]
    \includegraphics[width=0.85\columnwidth,clip]{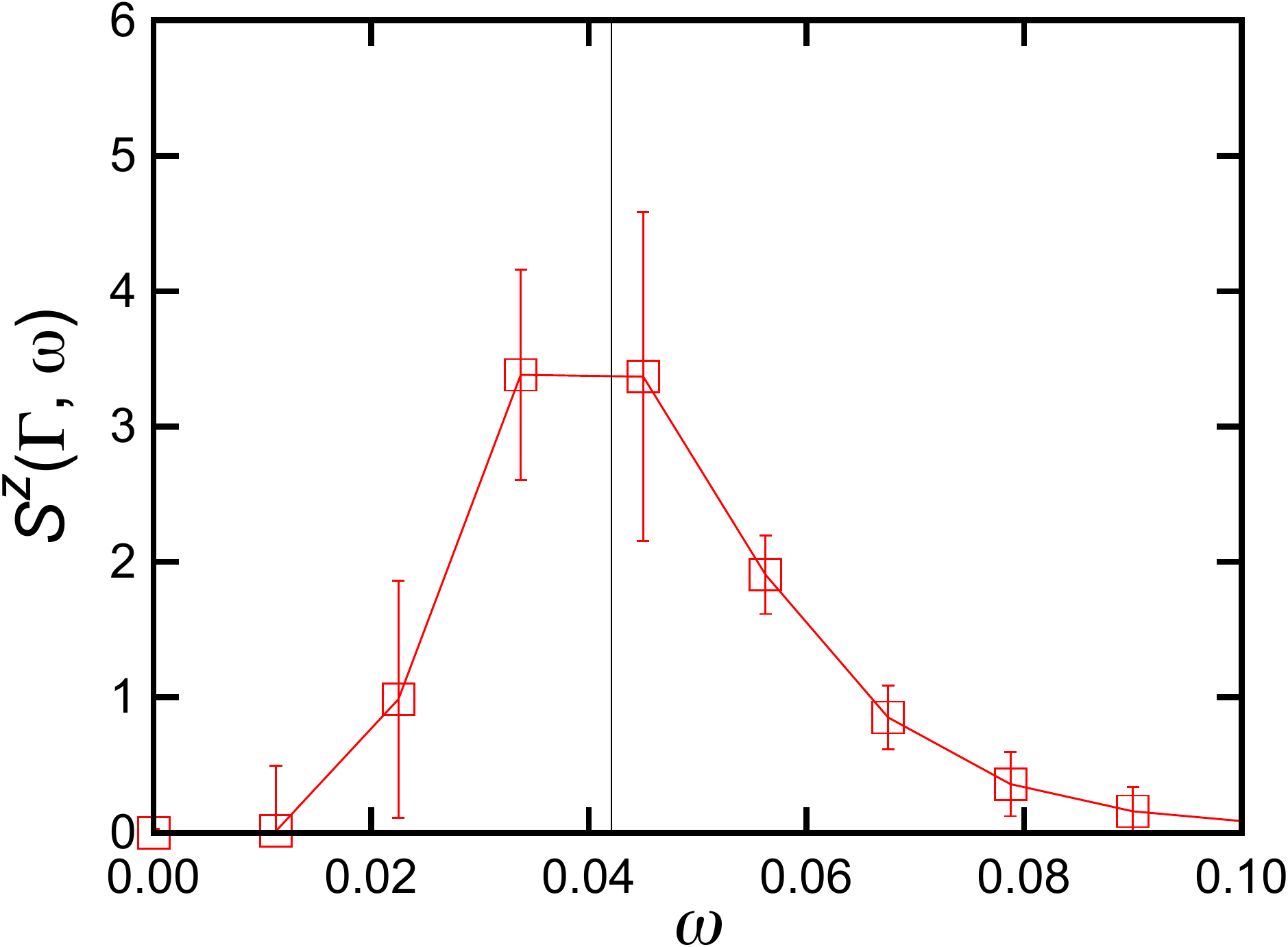}
    \caption{ \label{fig:fig16}
    $S^{z}(\Gamma,\omega)$ obtained by the Majorana CDMFT+CTQMC method for the FM case for $\alpha = 0.8$ at $T=0.003$. Vertical line at $\omega \sim 0.042$ represents the value of flux gap of the ground state calculated exactly.
    }
\end{figure}

Next, we examine the accuracy of the MEM for the data at sufficiently low $T$ with the analytical solution in the ground state~\cite{Knolle2014}. 
Figure~\ref{fig:fig16} shows $S^{z}(\Gamma,\omega)$ obtained by the Majorana CDMFT+CTQMC method for the FM case with $\alpha = 0.8$ at $T=0.003$.
In the ground state, the energy required to flip a single $\eta_r$ is $\Delta\simeq 0.042$ at $\alpha = 0.8$.  
Reflecting the flux gap, $S^{z}(\Gamma,\omega)$ at low $T$ has a $\delta$-function like peak at $\Delta\simeq 0.042$~\cite{Knolle2014}.
As shown in Fig.~\ref{fig:fig16}, our CDMFT+CTQMC result shows a peak at this energy, which is considered to precisely reproduce the low-energy structure of the dynamical spin structure factor.

From these observations, we consider that the MEM results for $S({\bf q},\omega)$ and $1/T_1$ in Sec.~\ref{subsec:DSF} and \ref{subsec:RelaxationRate} are accurate enough to discuss the $T$ and $\omega$ dependences.

\section{Spin correlations as functions of $T$ and $\omega$\label{app:S(w,T)}}

\begin{figure}[t]
    \includegraphics[width=\columnwidth,clip]{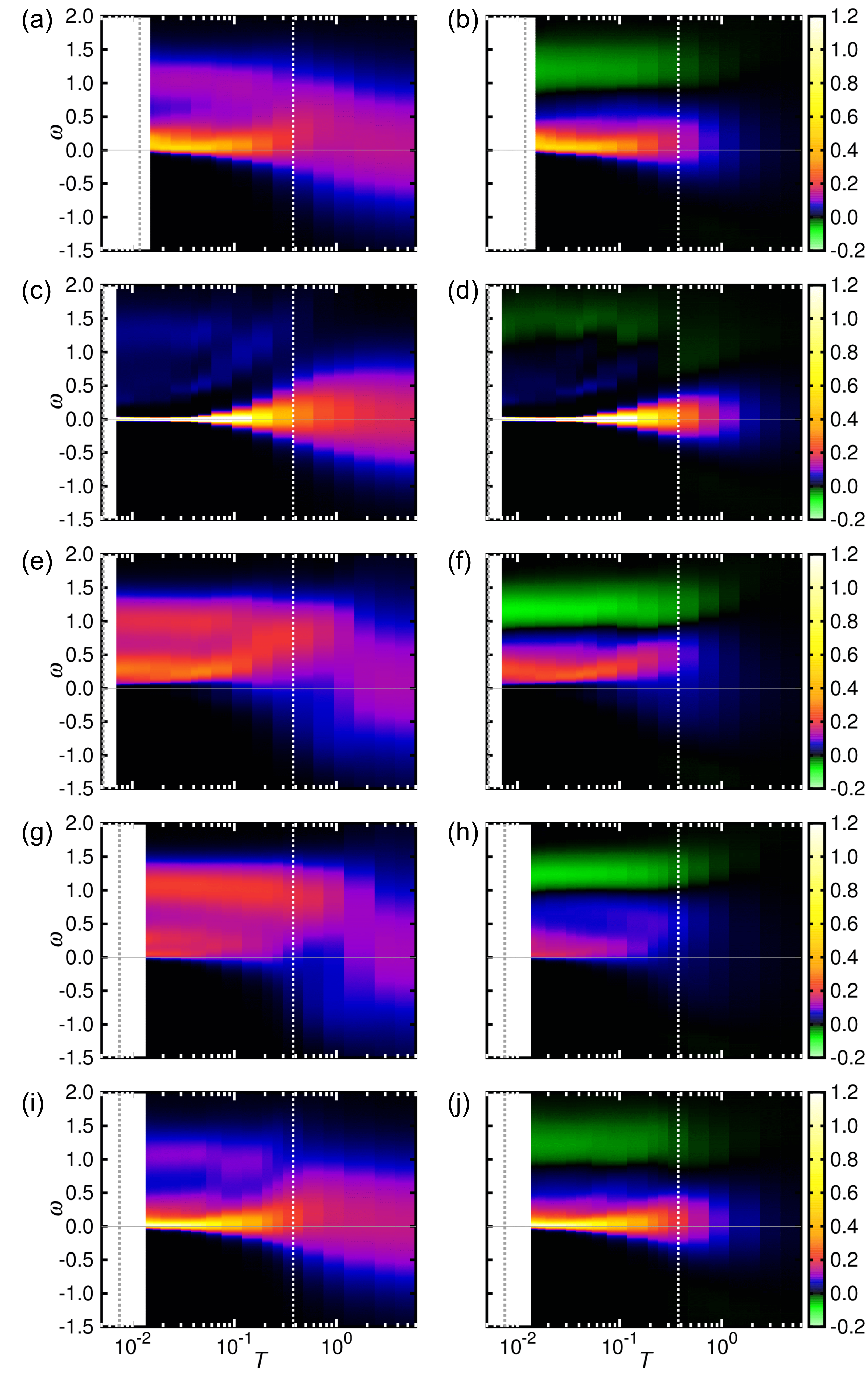}
    \caption{ \label{fig:fig17}
    Spin correlations as functions of $T$ and $\omega$ for the FM case: (a)(c)(e)(g)(i) onsite components $S_{j,j}(\omega)$ and (b)(d)(f)(h)(j) NN-site components $S_{\rm NN}(\omega)$. 
    (a) and (b) are for $S_{j,j}^x(\omega) = S_{j,j}^z(\omega)$ and $S_{\rm NN}^x(\omega) = S_{\rm NN}^z(\omega)$, respectively, at $\alpha=1.0$. 
    (c), (d), (e), and (f) [(g), (h), (i), and (j)] are for $S_{j,j}^x(\omega)$, $S_{\rm NN}^z(\omega)$, $S_{j,j}^z(\omega)$, and $S_{\rm NN}^z(\omega)$, respectively, at $\alpha=0.8$ ($1.2$).  
    The white and gray dotted lines indicate $T_{\rm H}$ and $T_{\rm L}$, respectively, for each $\alpha$. 
    }
\end{figure}

In this Appendix, we present the spin correlations as functions of $T$ and $\omega$, which are obtained by the MEM. 
Figure~\ref{fig:fig17} shows the results for onsite and NN-site components for $\alpha=1.0$, $0.8$, and $1.2$. 
The data are used to obtain the dynamical quantities in Sec.~\ref{subsec:DSF} and \ref{subsec:RelaxationRate}. 
\newline

\section{$T$ dependence of the Korringa ratio\label{app:Korringa}}

\begin{figure}[t]
    \includegraphics[width=0.8\columnwidth,clip]{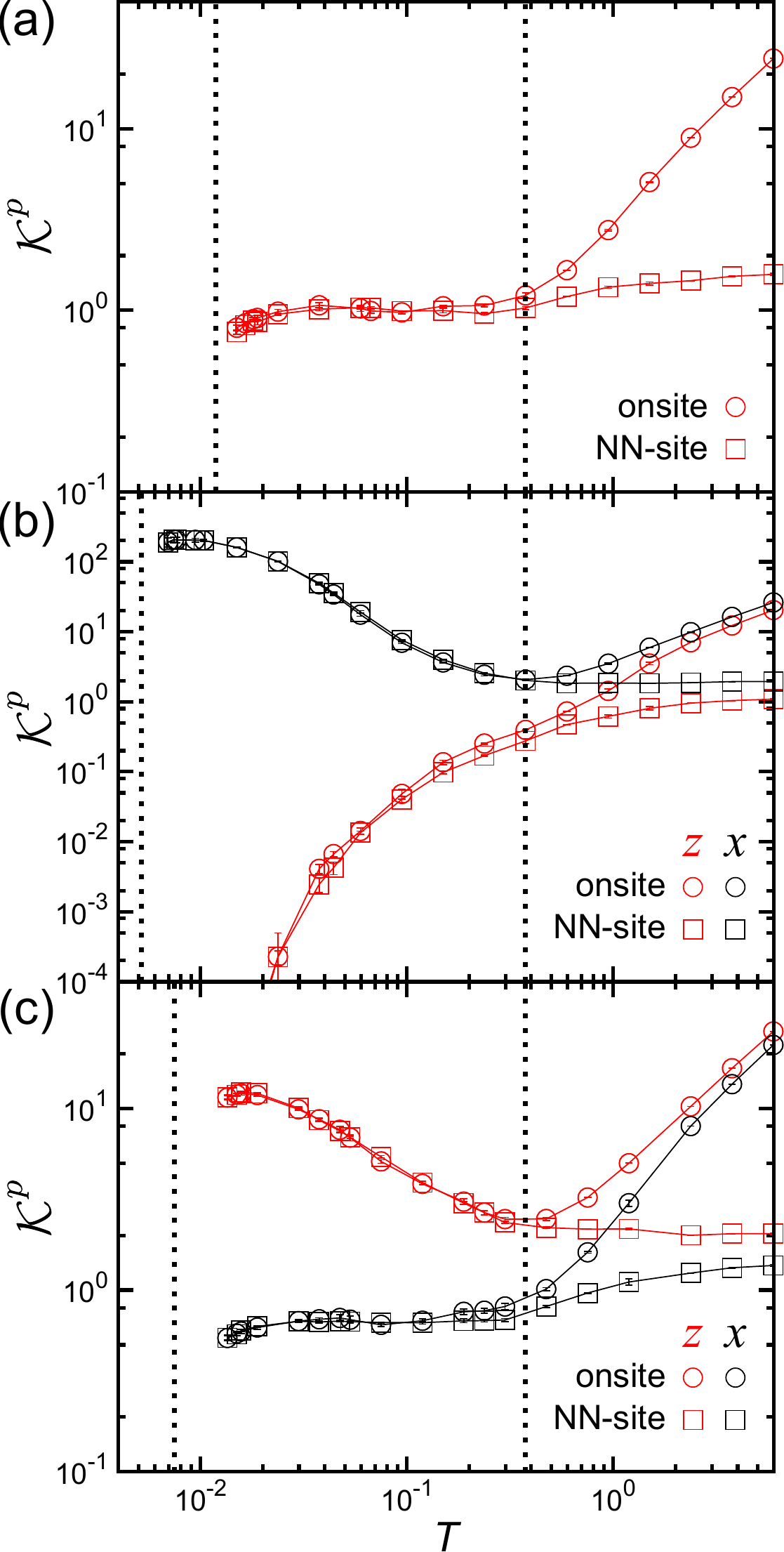}
    \caption{ \label{fig:fig18}
    $T$ dependences of the Korringa ratio $\mathcal{K}^p = 1/(T_1^pT (\chi^p)^2)$ for the FM case at (a) $\alpha=1.0$, (b) $\alpha=0.8$, and (c) $\alpha=1.2$ ($p=z,x$). 
    Note that $\mathcal{K}^z=\mathcal{K}^x$ for $\alpha=1.0$ and $\mathcal{K}^x=\mathcal{K}^y$ for all the cases from the symmetry. 
    The vertical dotted lines indicates $T_{\rm L}$ and $T_{\rm H}$ for each $\alpha$. 
    }
\end{figure}

\begin{figure}[t]
    \includegraphics[width=0.8\columnwidth,clip]{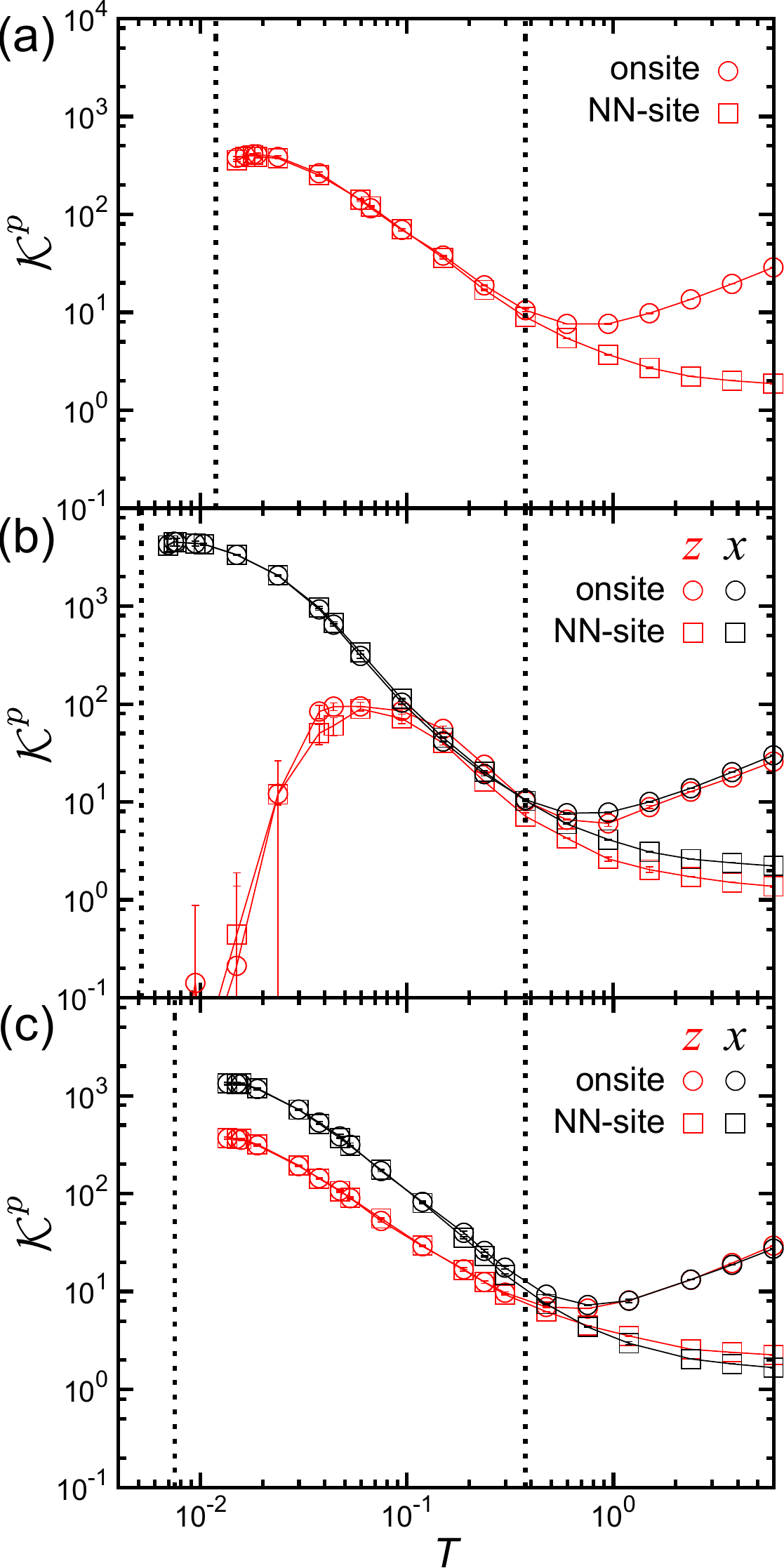}
    \caption{ \label{fig:fig19}
    $T$ dependences of the Korringa ratio $\mathcal{K}^p = 1/(T_1^pT (\chi^p)^2)$ for the AFM case at (a) $\alpha=1.0$, (b) $\alpha=0.8$, and (c) $\alpha=1.2$ ($p=z,x$). 
    The notations are common to those in Fig.~\ref{fig:fig18}. 
    }
\end{figure}

Figures~\ref{fig:fig18} and \ref{fig:fig19} display the $T$ dependences of the Korringa ratio defined as 
\begin{align}
\mathcal{K}^p = \frac{1}{T_1^pT (\chi^p)^2},
\end{align}
which is computed by using the NMR relaxation rate $1/T_1^p$ and the magnetic susceptibility $\chi^p$ obtained in Sec.~\ref{subsec:RelaxationRate} and \ref{subsec:Susceptibility}. 
Interestingly, as shown in Fig.~\ref{fig:fig18}(a), $\mathcal{K}^p$ for the isotropic FM case is almost constant close to $1$ for $T_{\rm L} \lesssim T \lesssim T_{\rm H}$, which is apparently consistent with the behavior expected for free electron systems. 
This is also the case for the $x$ component for the FM case with $\alpha=1.2$, as shown in Fig.~\ref{fig:fig18}(c). 
However, the suggestive behavior is presumably superficial, as the results for the AFM cases as well as for $\alpha=0.8$ behave differently with substantial $T$ dependence.

\begin{acknowledgments}
The authors thank M. Imada, Y. Kamiya, K. Ohgushi, S. Takagi, M. Udagawa, and Y. Yamaji for fruitful discussions.
Y. M. thanks A. Banerjee, C. D. Batista, K.-Y. Choi, S. Ji, S. Naglar, and J.-H. Park for constructive suggestions. 
This research was supported by Grants-in-Aid for Scientific Research under Grants No.~JP15K13533, No.~JP16K17747, and No.~JP16H02206. 
Parts of the numerical calculations were performed in the supercomputing systems in ISSP, the University of Tokyo. 
\end{acknowledgments}


\begin{thebibliography}{999}
\bibitem{note} 
$S^{p}_{\rm{NN}}(\omega)$ for the FM case ($S^{p}_{\rm{NN}}(\omega)_{\rm{FM}}$) and the AFM case ($S^{p}_{\rm{NN}}(\omega)_{\rm{AFM}}$) satisfy the relation $S^{p}_{\rm{NN}}(\omega)_{\rm{FM}} = -S^{p}_{\rm{NN}}(\omega)_{\rm{AFM}}$. Nevertheless, $S^{p}_{\rm{NN}}(\omega)_{\rm{FM}}$ and $S^{p}_{\rm{NN}}(\omega)_{\rm{AFM}}$ obtained by the present numerical procedure violates the relation because of the precision of the MEM. In the data presented in Sec.~\ref{sec:Result}, we calculate $S^{p}_{\rm{NN}}(\omega)_{\rm{FM}}$ by the procedure, and $S^{p}_{j,j'}(\omega)_{\rm{AFM}}$ are set as $S^{p}_{j,j'}(\omega)_{\rm{AFM}} = -S^{p}_{j,j'}(\omega)_{\rm{FM}}$.
\bibitem{note1} $\chi^x$ for the AFM case at $\alpha=0.8$ also shows similar behavior, but it is qualitatively explained by the Curie-Weiss law in Eq.~(\ref{eq:chi_CW}) for this case.
\bibitem{note2} Thermal averages $\langle\eta_{r_0}\rangle$ obtained by the CTQMC calculations deviate from $1$ even in the uniform case, although we fix $P(\eta_{r_0}=1)=1$ in the CDMFT. 
This is because the statistical weight of $\eta_{r_0}=-1$ is nonzero in the two-site impurity problem in the CTQMC calculations. 
Such a discrepancy of $\langle\eta_{r_0}\rangle$ between the CDMFT and CTQMC solutions never occurs when we do not fix $P(\eta_{r_0})$ by hand.
\end{thebibliography}
\end{document}